\documentclass{sig-alternate}
\usepackage{mathptmx} 

\newcommand{\ignore}[1]{}
\usepackage[pass]{geometry}
\usepackage{fancyhdr}
\usepackage[normalem]{ulem}
\usepackage[hyphens]{url}
\usepackage{hyperref}
\usepackage{color}
\usepackage{soul}

\usepackage{microtype} 

\fancypagestyle{firstpage}{
  \pagenumbering{arabic}
}

\usepackage{subfig}

\newcommand{\redHL}[1]{\textcolor{red}{#1}}

\newcommand{\hlnew}[1]{\textcolor{black}{#1}}

\newcommand{\shrink}{Eliminate vertical white-space}
\newcommand{\vshrink}[1]{
  \ifdefined\shrink 
	\vspace{-#1cm}
  \else
	\vspace{0cm}
  \fi
}

\newcommand{\arch}{GeNVoM}
\newcommand{\ssq}{{\em query}}

\newcommand{\srg}{{\em reference}}
\newcommand{\seed}{{\em seed}}

\title{
  Read Mapping Near Non-Volatile Memory  
} 
\author{S. Karen Khatamifard* \quad Zamshed Chowdhury* \quad Nakul Pande* \quad Meisam Razaviyayn$^\dagger$ \\[0.2cm] Chris Kim* \quad Ulya R. Karpuzcu*\\[0.5cm]
* University of Minnesota~~~~~~~~$^\dagger$ ~University of Southern California\\[0.5cm]
* \{khatami, chowh005, nakul, chriskim, ukarpuzc\}@umn.edu \quad $^\dagger$ razaviya@usc.edu
}

\begin{document}
\maketitle
\thispagestyle{firstpage}
\pagestyle{plain}

\noindent
\begin{abstract}

DNA sequencing is the physical/biochemical process of identifying the location
of the four bases (Adenine, Guanine, Cytosine, Thymine) in a DNA strand. As
semiconductor technology revolutionized computing, modern DNA sequencing
technology (termed Next Generation Sequencing, NGS) revolutionized genomic
research. As a result, modern NGS platforms can sequence hundreds of millions of
short DNA fragments in parallel. 
The sequenced DNA fragments, representing the output of NGS platforms, are
termed {\em read}s. Besides genomic variations, NGS imperfections induce noise
in {\em reads}. Mapping each {\em read} to (the most similar portion of) a
reference genome of the same species, i.e., {\em read mapping}, is  a common
critical first step in a 
diverse set of emerging bioinformatics applications.
%
%
Mapping represents a search-heavy memory-intensive similarity matching problem, therefore, can greatly benefit from
near-memory processing.
Intuition suggests using fast associative search 
enabled by Ternary Content Addressable Memory (TCAM) by
construction. 
However, the
excessive energy consumption and lack of support for similarity matching (under
NGS and genomic variation induced noise) renders direct application of TCAM infeasible, irrespective of volatility,
where only {\em non-volatile} TCAM can accommodate the large memory footprint in an
area-efficient way.
This paper introduces \arch, a scalable,
energy-efficient and  
high-throughput solution.
Instead of optimizing an
algorithm developed for general-purpose computers or GPUs, \arch\ rethinks the
algorithm and non-volatile TCAM-based accelerator design together from the ground up. 
Thereby \arch\ 
can improve 
the throughput 
by up to {113.5$\times$} ({3.6$\times$}); 
the energy consumption, by up to {210.9$\times$} ({1.36$\times$}), when compared to a
GPU (accelerator) baseline, which represents
one of the highest-throughput implementations known.


\end{abstract}

\section{Introduction}
\label{sec:intro}
\noindent 
DNA {sequencing} is the physical or biochemical process of extracting the order
of the four bases (Adenine, Guanine, Cytosine, Thymine) in a DNA strand.  As
semiconductor technology revolutionized computing, DNA sequencing technology,
termed {\em High-throughput Sequencing} or {\em Next Generation Sequencing}
(NGS), revolutionized genomic research.  As a result, modern NGS platforms can sequence
hundreds of millions of short DNA fragments in parallel.  
The sequenced fragments (which represent the NGS output) are referred to as
short {\em read}s 
and typically contain 100-200 bases~\cite{illumina}.  The focus of this paper is
{\em read mapping}, a common critical first step spanning a rich and diverse set
of emerging bioinformatics applications: mapping each 
NGS {\em read} to (the most similar portion of) a reference genome of the same species (which itself
represents a full-fledged assembly of already processed {\em read}s). 

\begin{figure}[htp] 
\vshrink{0.2}
\begin{center}
\includegraphics[width=0.75\columnwidth]{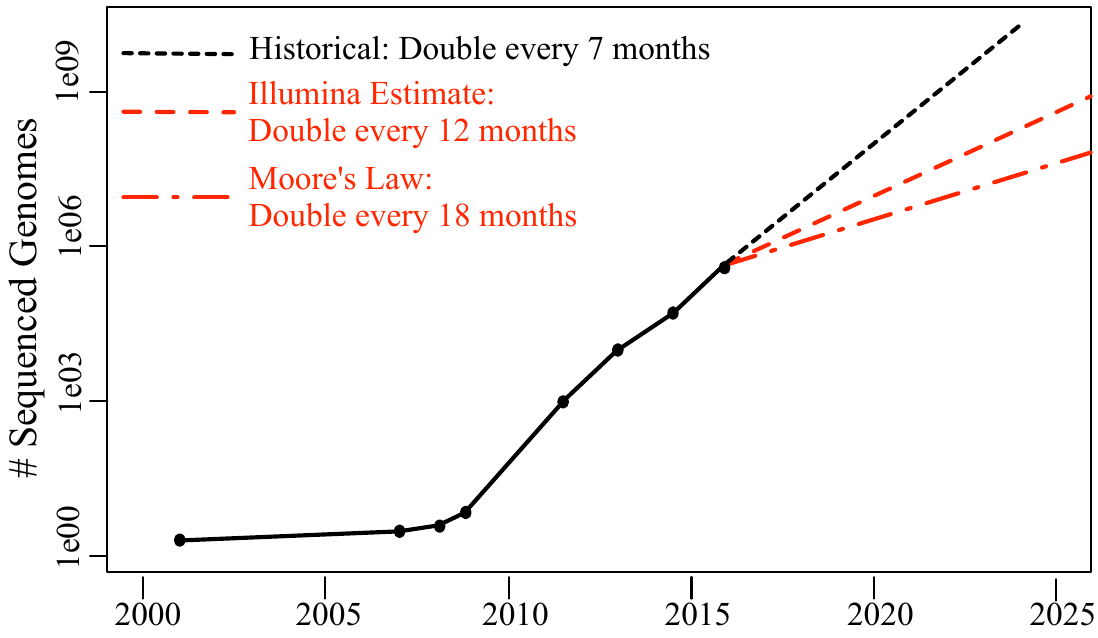} 
\vshrink{0.5}
\end{center}
\caption{Scaling trend for DNA sequencing~\cite{bigDataGenomics}.}
\label{fig:motiv} 
\vshrink{0.3}
\end{figure}

As a representative example, modern NGS machines from Illumina~\cite{illumina},
a prominent NGS platform producer, can sequence more than 600Giga-bases (Gba)
per one run, 200$\times$ the length of a human genome of approximately 3Gba,
which translates into hundreds of millions of output {\em reads}.
Fig.~\ref{fig:motiv}
depicts the scaling trend 
in terms of the total number of
human genomes sequenced.  \ignore{ (left y-axis), and sequencing capacity in
base-pairs (bps) per year (right y-axis).  } The values until 2015 reflect
historical publication records, with milestones explicitly marked. The values
beyond 2015 reflect three different projections: the first, following the
historical growth until 2015; the second, a more conservative prediction from
Illumina; the third, Moore's Law.  Historically, the total quantity 
has been doubling 
approx. 
every 7 months.  
Even the more
conservative projections from Fig.~\ref{fig:motiv} (i.e., 2$\times$ increase every 12 or 18
months) result in a very rapid
growth, which
challenges the throughput performance of {\em read mapping}.

The wildly increasing scale of the problem per Fig.~\ref{fig:motiv}
renders  well-studied pair-wise 
similarity detection 
algorithms inefficient~\cite{reviewAcc}.  Worse, {\em read}s are
subject to noise due to imperfections in NGS platforms and genomic variations, 
which adds to the complexity.
%
Both algorithmic solutions and hardware acceleration via GPUs~\cite{barraCUDA}
or FPGAs~\cite{hybridAccFPGA} therefore have to trade mapping accuracy
for throughput performance.
In other words, {\em read mapping} by definition is after {\em similarity
matching}. 
As optimizations are usually
confined to compute-intensive stages of mapping, 
considering 
scaling
projections from Fig.~\ref{fig:motiv}, most of these solutions are fundamentally
limited by data transfer overheads.  {
  In this paper, 
  we instead take a data-centric
  position to guide the design (and explore the design space) of scalable,
energy-efficient and high-throughput {\em read mapping}.  Specifically, instead of
optimizing an algorithm developed for general-purpose computers or GPUs, we
rethink the algorithm from the ground up along with the accelerator design.}

{\em Read mapping} represents a search-heavy memory intensive operation and
barely requires complex floating point arithmetic, therefore, can greatly
benefit from 
near-memory 
processing.  Intuition suggests using fast parallel
associative search, enabled by Ternary Content Addressable Memory (TCAM) by
construction, in matching short {\em read} patterns with
portions of the large reference genome (stored in TCAM). 
As we will explain in Section~\ref{sec:biocamArch}, however, 
only {\em non-volatile}
TCAM can accommodate the large
memory footprint in an area- and energy-efficient manner~\cite{ipekTCAM}.
%
Even then, brute-force non-volatile TCAM search over as large of a search space as {\em read
mapping} demands induces excessive energy consumption, rendering (non-volatile)
TCAM-based acceleration infeasible. At the same time, by construction,
(non-volatile) TCAM cannot handle similarity matching under NGS or genomic
variation induced noise. 

This paper provides an effective solution, \arch, to tap the potential of {\em
non-volatile}
TCAM for scalable, energy-efficient high-throughput
{\em read mapping}. \arch\ 
\begin{list}{\labelitemi}{\leftmargin=1em}
\vshrink{0.2}
  \itemsep-0.2em 
  \item introduces a novel similarity matching mechanism 
	{for resistive {non-volatile} TCAM} (which can only handle exact matches by
	construction),
    to trade mapping accuracy for
	throughput and energy efficiency in a much more scalable manner than
	existing solutions;
  \item features a novel 
	{genomic} data representation 
	for efficient similarity
	matching without compromising storage complexity;
  \item tailors common search space pruning approaches	
	to its novel similarity
	matching mechanism in order to identify and discard unnecessary
	non-volatile TCAM
	accesses (and thereby, to 
	prevent
	excessive energy consumption); 
    \item accounts for 
	  the most  prevalent manifestations of noise induced by NGS
	  imperfections and genomic variations during similarity matching, including
	  (base) gaps and
	  insertions/deletions in {\em reads};  
%
  \item 
	employs multi-phase hierarchical similarity matching
	to enhance mapping
	accuracy and scalability, where each
	{phase performs progressively more
	sophisticated mapping, considering only the subset of {\em read}s the
	previous phase fails to map}. 
\vshrink{0.2} 
\end{list}
\ignore{
While enabling approximate or similar matches in TCAMs have been explored
before, we discuss in Section~\ref{sec:rel} why such schemes are not
applicable considering the scale of the problem.
}
In the following, we introduce a proof-of-concept \arch\ implementation.
Specifically, 
Section~\ref{sec:biocamArch} discusses basics;
Section~\ref{sec:pract} covers implementation details; 
Sections~\ref{sec:setup} and~\ref{sec:eval} detail the evaluation;
Section~\ref{sec:rel} provides a compare and contrast to related work; 
and
Section~\ref{sec:conc} summarizes our findings.


\ignore{
This paper provides an effective solution to tap the potential of non-volatile
TCAM for scalable, energy-efficient high-throughput
{\em read mapping}, \arch, which 
\begin{list}{\labelitemi}{\leftmargin=1em}
\vshrink{0.2}
  \itemsep-0.3em 
  \item features a light-weight filtering mechanism to identify and discard unnecessary memory
accesses, and thereby, to effectively prune the search space and prevent
excessive energy consumption; 
  \item introduces a novel similarity match mechanism to trade mapping accuracy for
throughput and energy efficiency in a much more scalable manner than existing solutions;
  \item is expandable to similar search-intensive problems from other
	application domains beyond bioinformatics.
\vshrink{0.2} 
\end{list}
}

\section{\arch: Macroscopic View}
\label{sec:biocamArch}
\ignore{
\noindent  Short (i.e., 100-200 base long) {\em read}s from modern Illumina NGS
platforms~\cite{illumina} constitute more than 90\% of all {\em read}s in the
world currently.
This dominance is unlikely to quickly change in the near future due to the
progressively dropping sequencing cost of short {\em read} technologies,
rendering them significantly more cost-efficient than the long {\em read}
counterparts such as PacBio~\cite{pacBio} or Oxford
Nanopore~\cite{branton2008potential} (where {\em read} lengths can exceed tens
of thousands of bases).  The key benefit of long {\em read} sequencing
technologies comes from the capability of directly extracting long-range
information, and not necessarily from higher accuracy.  That said, many emerging
recent technologies such as 10xGENOMICS~\cite{10xG} can 
obtain long-range information from short {\em read}s.  Although it is very hard
to predict the future exactly, 
considering practical facts such as market share and market caps on top, we
believe that short {\em read} platforms will remain prevalent
at least in the near future. Accordingly, \arch\ is designed for short {\em
read} mapping.
}

\vshrink{-.1}
\noindent  
{\bf Scope:} Short (i.e., 100-200 base long) {\em read}s from modern Illumina NGS
platforms~\cite{illumina} constitute more than 90\% of all {\em read}s in the
world currently.  Accordingly, \arch\ is designed for short {\em read} mapping.

\vshrink{-.1}
\noindent {\bf Terminology:} Without loss of generality, 
we will refer to each {\em read} simply as a \ssq; and the reference
genome, as the \srg. 
Each \ssq\ and the \srg\ represent strings of characters from the alphabet \{A,
G, C, T\} which stand for the bases \{Adenine, Guanine, Cytosine, Thymine\}. 
The inputs to \arch\ are 
a dataset of
{\ssq}s and the
{\srg}, where the \srg\ is many orders of magnitude longer than each \ssq. For
example, if the \srg\ is the human genome, \srg\ length is
approximately 3$\times 10^9$ bases.
On the other hand, technological capabilities of modern NGS platforms limit the
maximum \ssq\ length. 

\begin{figure}[tp]
\vshrink{0.2}
\begin{center}
  \includegraphics[width=\columnwidth]{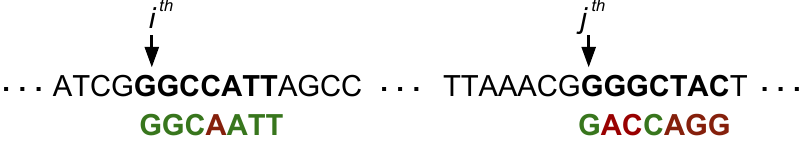} 
  \vshrink{0.7}
  \caption{Read mapping example.
  \label{fig:problem}}
\end{center}
\vshrink{.8}
\end{figure}

\subsection{Problem Definition: Read Mapping}
\label{sec:pr}

\vshrink{-.1}
\noindent {\bf Basics:} {\em Read mapping} entails finding the most {\em similar} portions of
a given {\em reference} to each {\em query} from a dataset corresponding to the
same species, as output by an NGS machine.  Fig.~\ref{fig:problem} demonstrates
an example, with different portions from the same {\em reference} on top; two
sample {\em query}s to be mapped, at the bottom. The first (second) {\em query}
results in one (five) base-mismatch(es) when aligned to the $i^{th}$ ($j^{th}$)
base of the {\em reference}. The {\em query} length is not representative, but
simplifies demonstration.

\arch's input {\ssq}s are subject to noise due to imperfections in NGS platforms and
potential genomic variations. 
Therefore, {\em read mapping} by definition is after {\em similarity} rather
than an {\em exact match}.
Hence, for each input \ssq, \arch\ tries to locate the {\em most similar} sub-sequence
of the \srg\ to the \ssq, and
returns the range of its indices. 

\vshrink{-.1}
\noindent 
{\bf Mapping Reverse Complement of {\em read}s:} 
It is not uncommon for NGS platforms to sequence DNA strands in reverse
direction. This happens when sequencing starts from the last base of a DNA
strand. 
In this case,  
the platform
outputs the reverse complement of a {\em read}
by interchanging A with T, and C with G. For example, 
the reverse complement of the sequence ACCGCCTA is TAGGCGGT.
NGS platforms typically sequence almost half of the DNA strands in reverse order, 
hence, \arch\ is designed to 
handle reverse complements.

\ignore{
It is not uncommon
for NGS platforms to sequence DNA strands in reverse direction. This happens
when sequencing starts from the 
last base of a DNA strand. In this case, NGS platforms output the
complement of each base, by interchanging A with T, and C with G. 
For example, the reverse complement of the sequence ACCGCCTA 
is given by TAGGCGGT. Since
NGS platforms typically sequence almost half of DNA strands in reverse order, 
\arch\ is designed to 
handle reverse complement of 
{\em read}s (i.e., {\em query}s), as well.
}

\vshrink{-.1}
\noindent
{\bf Sources of Noise in Similarity Matching:}
{
In general, the sequenced genome (where the {\em reads} are coming from) is
  expected to be slightly different from the {\em reference} genome, even though
  they represent the very same species. 
{\em Genomic variations} induce such differences, which can lead to
base-mismatches between the {\em query}s and the {\em reference}, 
since the {\em query}s come from the sequenced genome as opposed to the {\em
reference}.
{NGS platform} imperfections, as well, can result in false base-mismatches between the {\em
query}s and the {\em reference}, due to the so-called {\em read errors} during
sequencing. We will next discuss the most prevalent manifestations of genomic
variations and read errors.
}

\vshrink{-.1}
\noindent
{\bf Noise Manifestation:}
{
Most common genomic variations and read errors 
manifest themselves in three ways:
Random {\em insertion} of a base, random {\em deletion} of a base, and random {\em
substitution} of a base with another.
Insertions and deletions are often referred to as {\em
indels}.
}
%
The expected rates of indels and substitutions depend on the type of
genomes (hence species) and the NGS technology.

\begin{figure}[tp]
\vshrink{0.9}
\begin{center}
  \includegraphics[width=.9\columnwidth]{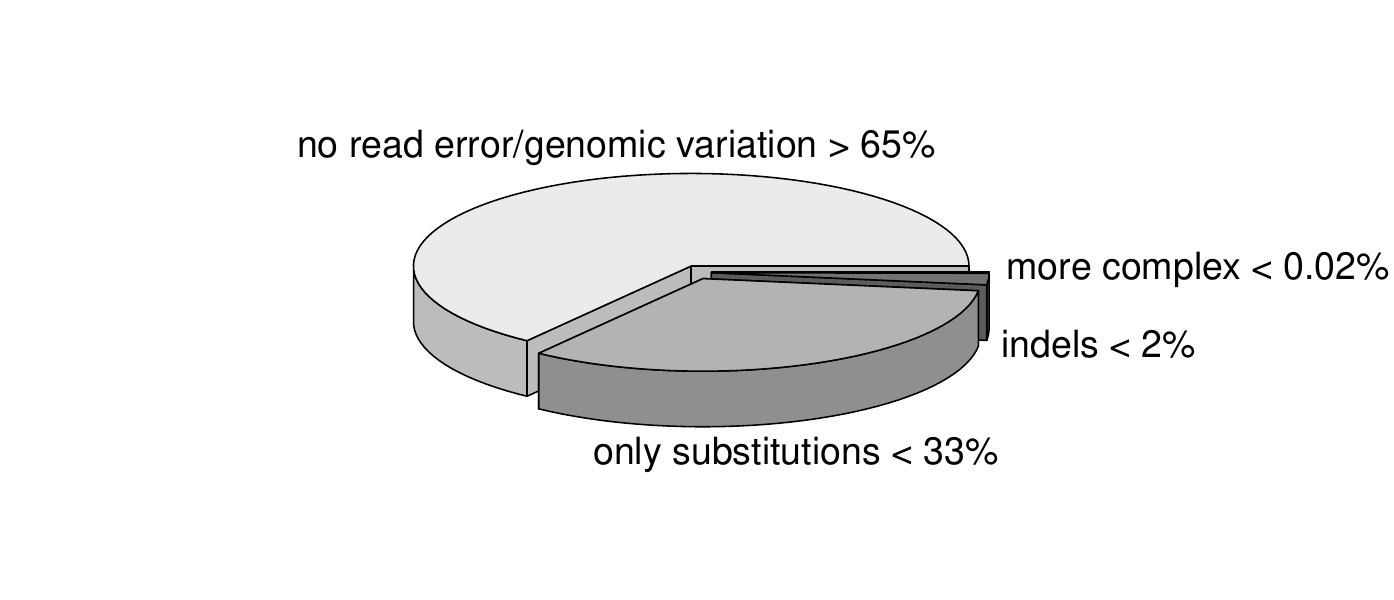} 
  \vshrink{.8}
  \caption{
	Manifestation of genomic variations \& read errors.
  \label{fig:pie}}
\end{center}
\vshrink{0.8}
\end{figure}

As a representative example, Fig.~\ref{fig:pie}
demonstrates the typical 
share of 
{\em read}s  
having no read errors/genomic
variations (>65\%), at least one substitution (and no indels) (<33\%), at least
one indel (<2\%), 
and more complex manifestations (<0.02\%)~\cite{insdel,var1,var2,var3}.
{Mapping under rare complex manifestations (such as long-indels, gaps, base
  duplications or inversions) is a daunting task, and
to date there is no widely-accepted algorithm to cover all~\cite{var2,var6}.
As Fig.~\ref{fig:pie}
shows, substitutions are dominant.
Although indel rate is on the lower side, detecting indels is critical for many
downstream applications. 
Hence,
\arch\ is designed to cover both substitutions and indels, but optimized for the
common case (no read error/genomic variation and only substitutions), which
covers more than 98\% of the {\em read}s per Fig.~\ref{fig:pie}. As we will
demonstrate in Section~\ref{sec:hier}, straight-forward expansion of \arch\ to
more complex manifestations such as gaps is also possible.

\ignore{
To summarize, {\em read mapping} is a search-heavy 
similarity
matching operation and can
greatly benefit from parallel in- or near-memory associative search enabled by
Ternary Content Addressable Memory (TCAM). We will next discuss the feasibility
of TCAM-based acceleration for {\em read mapping}.
}

\ignore{
\arch\ is an accelerator for short {\em read mapping}, where typical {\em read}s
are 100-200 bases long. Fundamentally, \arch\ hardware can support mapping for
much longer {\em read}s. What makes \arch\ an accelerator for short {\em read}s
is not the actual length of the {\em read}s, but the types of errors \arch\ can
handle. This is because the prevalence and probabilities of errors represent a
strong function of the {\em read} length. The proof-of-concept \arch\ design
does not cover indels. 
According to Illumina data-sheets, machine-induced indels in short {\em read}s
are negligible~\cite{insdel} (which only become prominent for long {\em read}s).
On the other hand, while also rare, it is important to capture {\em
mutation-induced} short indels, since this information is usually what we are
after. 
Investing significant hardware resources to support mutation-induced short
indels, however, which are expected to affect only a very small number of {\em
read}s, is likely to render a sub-optimal design.  We instead envision a
hierarchical solution, which first maps {\em read}s by assuming  only
substitution errors (as the proof-of-concept \arch\ design does) and considers
short {\em indels} in a subsequent mapping step (likely in software) for
refinement. \redHL{In Section~\ref{sec:acc}, we will quantify \arch's mapping accuracy
in the presence of such mutation-induced short indels.} 
}

\begin{figure*}[htp]
  \begin{center}
  \includegraphics[width=0.75\textwidth]{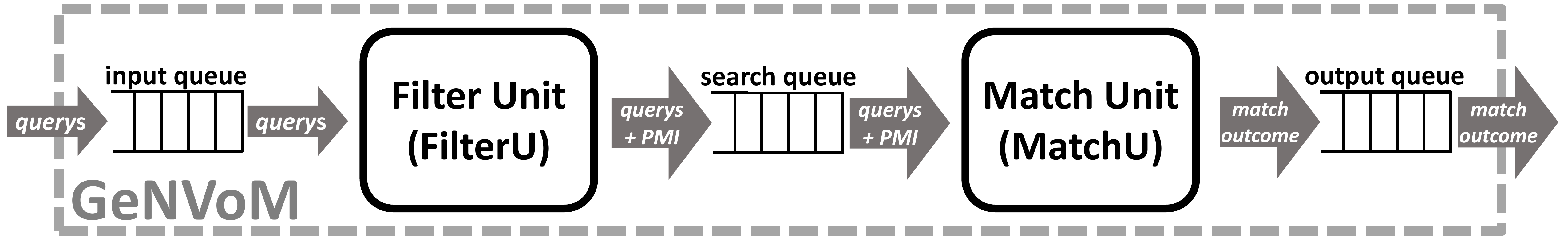}
\vshrink{0.3}
 \caption{Structural organization.
  \label{fig:biocam}}
\end{center}
\vshrink{0.7}
\end{figure*}

\subsection{{Why Naive (Non-Volatile) TCAM-based \\ Acceleration Does Not Work}}
\label{sec:motiv}
\ignore{
{\em Read mapping} essentially is a search-heavy 
similarity matching problem.
Therefore, facilitating fast parallel in-memory pattern matching, TCAM-based acceleration
is particularly suitable for {\em read mapping}. 
}

\noindent {\em Read mapping} essentially is a search-heavy memory intensive
pattern matching problem.  This suggests TCAM-based acceleration, which by
construction can support fast parallel in-memory search.
%
\ignore{
However, in matching short {\em read} patterns with
portions of the large reference genome (stored in TCAM), brute-force
non-volatile TCAM search over the large search space
induces excessive energy consumption, as we will demonstrate next.
That said, even if the energy consumption 
}
TCAM is a special variant of associative memory (which permits data retrieval by
indexing by content rather than by address) that can store and search the
``don't care'' state X in addition to a logic 0 or 1. {Considering the scale
of the problem}, however, only {\em non-volatile} TCAM can accommodate the large
memory footprint in an area- and energy-efficient manner~\cite{ipekTCAM}.
\ignore{
Moreover, 
non-volatile TCAM can overcome area and energy inefficiencies
of CMOS-based TCAM~\cite{ipekTCAM}.
}

We will next look into the energy consumption of {\em read mapping}, comparing a
non-volatile TCAM-based implementation with a highly optimized GPU-based
solution deploying one of the fastest known algorithms to date~\cite{luo2013soap3}.
The non-volatile TCAM  mimics the least energy-hungry implementation from Guo et
al.~\cite{ipekTCAM}, corresponding to an array size of 1K$\times$1Kbits =
1Mbits.  For this design point, searching for a pattern of length 1Kbits (which
represents the maximum-possible length, i.e., the row length) in the entire
array takes approximately 2.5ns and consumes 245nJ.

If we simply encode each base from the alphabet \{A, G, C, T\} using 2 bits, and
if a human genome of approximately 3Giga-bases (= 6Gbits) represents the \srg,
  the \srg\ can fit into 6K TCAM arrays (of size 1K$\times$1Kbits = 1Mbits).  
For each \ssq\ of a typical length of 100 bases~\cite{illumina}, i.e.,  200
bits, the following naive procedure can cover the entire search space: By
construction, each 1K$\times$1Kbit TCAM array can search for at most one 1Kbit
pattern  at a time, which resides in a query register.  We can align the most
significant bit of the (200bit-long) \ssq\ with the most significant bit
position of TCAM's 1Kbit query register, and pad the remaining (1K-200) bits by
Xs, for the very first search in the array.
We can then repeat the search by shifting the contents of TCAM's query register
(i.e., the padded \ssq) to the right  by one bit at a time, leaving the unused
more significant bit positions with Xs, until the least significant bit of the
{\em query} reaches the least significant bit position in the query register.
The total number of these bit-wise shifts (and hence, searches) would be in the
order of the row length $\approx$ 1K.  Putting it all together, mapping a given
\ssq\ to the \srg\ in this case would take around 1K searches in each of the 6K
arrays, with 245nJ consumed per search. The overall energy consumption therefore
would become 6K $\times$ 1K $\times$ 245nJ $\approx$ 1500mJ. 

The GPU solution from Luo et al.~\cite{luo2013soap3} on the other hand, can
process 133.3K {\ssq}s per second.
Hence, it takes 1/133.3K seconds to
map a single \ssq.  Even under the unrealistic assumption (favoring TCAM) that
the entire peak average power (TDP) goes to mapping a single \ssq\ to the \srg,
the energy consumption would become at most 235W $\times$ (1/133.3K)s $\approx$
1.8mJ.

The GPU and TCAM designs feature similar technology nodes, however, even 
by favoring TCAM, the TCAM-based naive implementation consumes approx. 3 orders of
magnitude more energy than the GPU-based. This
difference stems from the gap in the size of the search spaces. While the
TCAM-based design considers the entire search space to cover all possible
alignments, the GPU-based design first prunes the search space to eliminate
infeasible alignments, which in turn leads to orders of magnitude less number of
(search) operations.  
\arch, 
while deploying non-volatile TCAM arrays,
adopts a similar pruning strategy  
to enable more energy-efficient search. 

Even if excessive energy consumption was not the case, (non-volatile) TCAM has
another fundamental limitation which hinders applicability to {\em read
mapping}. As we will detail in Section~\ref{sec:similar}, even in the presence
of ``don't cares'', TCAM cannot handle similarity matching considering various
manifestations of noise due to NGS errors and genomic variations
(Section~\ref{sec:pr}).  

To summarize, both, {\em the excessive energy
consumption and 
%
lack of support for similarity matching render a direct adaption of non-volatile
TCAM-based search infeasible}. {The energy overhead of conventional
volatile
TCAM would be even higher}~\cite{ipekTCAM}, while the restriction on similarity matching
directly applies irrespective of volatility.


\arch\ 
unlocks the throughput potential of non-volatile TCAM in a scalable and
energy-efficient manner through 

\begin{list}{\labelitemi}{\leftmargin=1em}
\vshrink{0.2}
\itemsep-0.3em 
  \item a novel {\bf non-volatile} resistive TCAM design capable of {\bf similarity
  matching} (Sect.~\ref{sec:similar}); 
  \item a novel {\bf genomic data representation for similarity matching} without
	compromising storage complexity (Sect.~\ref{sec:encode});  
  \item a common filtering mechanism~\cite{filterBook} for {\bf search space
	pruning adapted to non-volatile
  similarity matching} to prevent excessive energy consumption (Sect.~\ref{sec:prune});
  \item {\bf hierarchical similarity matching} to maximize mapping accuracy without
	compromising scalability (Section~\ref{sec:hier}).
\vshrink{0.2} 
\end{list}

\ignore{
While {resistive non-volatile} 
TCAMs capable of similarity matching have been explored
before, Section~\ref{sec:rel} reveals why such schemes are not
applicable considering the scale of {\em read mapping}. 
}

Designed for similarity search (in the presence of NGS or genomic variation
triggered noise), \arch's {non-volatile} TCAM arrays can directly handle
substitutions, by construction (Sect.~\ref{sec:similar}).  {To cover} indels and
more complex corruptions, on the other hand, \arch\ adapts 
{\em anchoring} within its multi-phase mapping hierarchy (Sect.~\ref{sec:hier}). 
The key insight is that complex corruptions that can lead to failed mappings are
much less likely to occur in {\em all} portions of a {\em read} simultaneously.
This makes anchoring very effective -- a divide and conquer technique which
entails chunking the {\em read} (at an anchored base position) to shorter substrings
and attempting mapping on each chunk simultaneously. Thereby, problematic
chunk(s) (and hence the entire {\em read}) simply follow the alignment dictated
by the less-corrupted chunks, which by construction renders the most accurate
mapping under noise. 
Numerous prevalent {\em read mapping} algorithms~\cite{bowtie1} therefore use
anchoring-based techniques for complex corruptions. 

\ignore{
In order to identify and discard unnecessary
memory accesses, and thereby, to prevent excessive energy consumption during TCAM search, 
\arch\
tailors well-known search space pruning (i.e., filtering) techniques 
\redHL{[let's cite the original book here, to avoid darwin complains. I will get it from
Meisam.]}
to its novel
similarity matching mechanism.
While \redHL{resistive non-volatile} 
TCAMs capable of similarity matching have been explored
before, Section~\ref{sec:rel} reveals why such schemes are not
applicable considering the scale of {\em read mapping}. 
The key features of \arch\ are
\begin{list}{\labelitemi}{\leftmargin=1em}
  \item a novel, \redHL{resistive non-volatile} TCAM similarity matching mechanism,
   to trade mapping accuracy for
	throughput and energy efficiency without compromising scalability; 
  \item a novel \redHL{genomic} data representation to facilitate efficient similarity
	matching without compromising storage complexity;  
  \item 
	employing multi-phase hierarchical mapping,
	 to maximize mapping
	accuracy, where each
	\redHL{
	phase performs more
	sophisticated mapping, considering only the subset of {\em read}s the
	previous phase fails to map. 
	}
\end{list}
\redHL{[Why are we repeating the contributions here? in case we need more space, we can get rid of these.]}
}

\subsection{Hardware Organization}
\label{sec:biocam}
\label{sec:biocamorg}
\noindent 
Fig.~\ref{fig:biocam} provides the
structural organization.
\arch\ pipeline comprises two major units: Filter Unit (FilterU) and Match Unit
(MatchU).  Each \ssq\ from the dataset to be mapped streams into the (first
stage of the) \arch\ pipeline (i.e., FilterU) over the {\em input queue}.  Once
the mapping completes, the outcome streams out of the (last stage of the) \arch\
pipeline (i.e., MatchU) over the  {\em output queue}.  Non-volatile TCAM arrays
(which feature \arch's novel similarity matching mechanism) within MatchU keep the
entire \srg.

Input and output queues handle the communication to the outside world, by
retrieving {\ssq}s 
on the input end, and upon completion of the
mapping, by providing the indices of the most {\em similar} sub-sequences of the
\srg\ to each \ssq, on the output end.
%

FilterU filters (indices of) sub-sequences of the \srg\ which are
more likely to result in a match to the incoming \ssq, by examining
sub-sequences of the incoming \ssq\ itself. We call these indices {\em
potentially matching indices}, PMI.  {FilterU feeds the MatchU with a stream of
$<$PMI, {\em query}$>$ tuples over the {\em search queue}.} 
MatchU
in turn conducts the search by only considering PMI of the \srg. In this manner,
\arch\ prunes the search space.  

The input queue feeds the \arch\ pipeline with the {\ssq}s to be mapped to the
\srg.  The \ssq\ dataset resides in memory.  \arch\ initiates the streaming of
the {\ssq}s into the memory-mapped input queue over a  Direct Memory
Access (DMA) request.  The input queue in turn sends the {\ssq}s to FilterU for
search space pruning before the search takes place. 
Finally, for each \ssq, once the mapping completes, the output queue
collects from MatchU the indices of the sub-sequence of the \srg\ featuring the
most similar match to the \ssq. The output queue is memory-mapped, too. So,
\arch\ writes back these indices to a dedicated memory location, over DMA. 

In the following, we will detail the steps for \ssq\ processing in each unit in
case of a match.  If no sub-sequence of the \srg\ matches the input \ssq, 
no mapping takes place, and \arch\ updates a
dedicated flag at the memory address to hold the result.  \arch\ can detect such
failed mapping attempts during processing at FilterU or at MatchU.

\begin{figure}[tp]
\begin{center}
  \includegraphics[width=0.8\columnwidth]{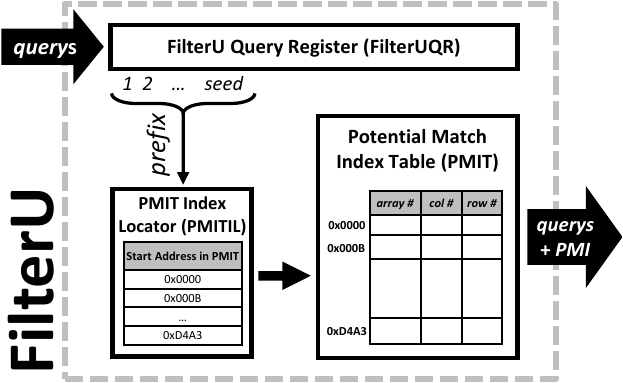}
\vshrink{0.3}
\caption{Filter Unit (FilterU).
  \label{fig:fu}}
\end{center}
\vshrink{.9}
\end{figure}

\subsubsection{Filter Unit (FilterU)} 
\noindent Fig.~\ref{fig:fu} provides the structural
organization of FilterU, which serves the compaction of the search space for
each \ssq\ to be mapped to the \srg, as follows: We will refer to each
sub-sequence of length \seed\ as a {\em prefix}, where \seed\ represents a
design parameter and assumes a much lower value than the \ssq\ length.  
As each {\em prefix} is a string of characters from the 4-character alphabet
\{A, G, C, T\}, a {\em prefix} of length \seed\ can take 4$^{seed}$ different
forms.  
Considering the size of the problem, the \srg\ is likely to occupy multiple TCAM arrays. 
FilterU relies on a pre-processing step which entails identifying
each
{\em prefix} of length \seed\ in the \srg, and recording the
TCAM array, column and row number of the corresponding occurrence.  {\em
Potential Match Index Table} PMIT keeps this information.  

However, as the same {\em prefix} may occur multiple times along the \srg\
string, PMIT may contain multiple entries for the very same {\em prefix}.
Therefore, FilterU has another table called {\em PMIT Index Locator} (PMITIL)
for bookkeeping. PMITIL serves as a dictionary of 4$^{seed}$ entries, considering all possible 4$^{seed}$ values
of the (\seed-long) {\em prefix}.  
{  
Each PMITIL}
  entry refers to a specific {\em prefix} value, and  keeps the start 
  address in
  PMIT where the TCAM indices for the
  corresponding occurrence of the {\em prefix} (along the {\em reference}) reside. 
  As PMIT is organized to keep multiple occurrences (along the {\em reference})
  of the same {\em prefix} consecutively,
  it suffices to keep per PMITIL entry just the start address (in the PMIT) for the
first occurrence. The end address in this case simply corresponds to the start
address stored in
the next PMITIL entry.
\ignore{
  \arch\ arranges PMIT to
  store multiple entries corresponding to the same prefix at consecutive
  addresses. 
  Therefore, it suffices to keep just the start address in PMITIL (in
  fact, the end address becomes the same as the start address for the next
  PMITIL entry).
}
  \ignore{
	  Each PMITIL
  entry refers to a specific {\em prefix} value, and  keeps the start address in
  PMIT where the matching TCAM indices for the
  corresponding {\em prefix} reside. \arch\ arranges PMIT to
  store multiple entries corresponding to the same prefix at consecutive
  addresses. Therefore, it suffices to keep just the start address in PMITIL (in
  fact, the end address becomes the same as the start address for the next
  PMITIL entry).
  }

  PMIT and PMITIL generation constitutes a pre-processing step which \arch\
needs to perform only once, offline, for each \srg, before {\em read mapping}
starts. As {\em read mapping} entails mapping a large number short {\em read}s
to a given {\em reference} of the same species, this overhead does not apply to
runtime, and is easy to amortize. 

Upon receipt of a new \ssq\ from the head of the input queue, FilterU uses the
first \seed\ bases of the \ssq\ as the {\em prefix} to consult PMITIL, and
subsequently, PMIT.  FilterU keeps the \ssq\ being processed in the FilterU {\em
Query Register} (FilterUQR) as filtering is in progress.  If there is a match in
the PMI tables,
FilterU 
first broadcasts the \ssq\ being processed to all TCAM arrays. Then,
it sends the corresponding TCAM array, column and row number (i.e., the
Potential Match Indices, PMIs) to
MatchU, over the {\em search queue}. 
We will
refer to these TCAM coordinates as $array_\#$, $col_\#$, and $row_\#$,
respectively.

\begin{figure}[tp]
 \vshrink{0.2}
  \begin{center}
  \includegraphics[width=0.7\columnwidth]{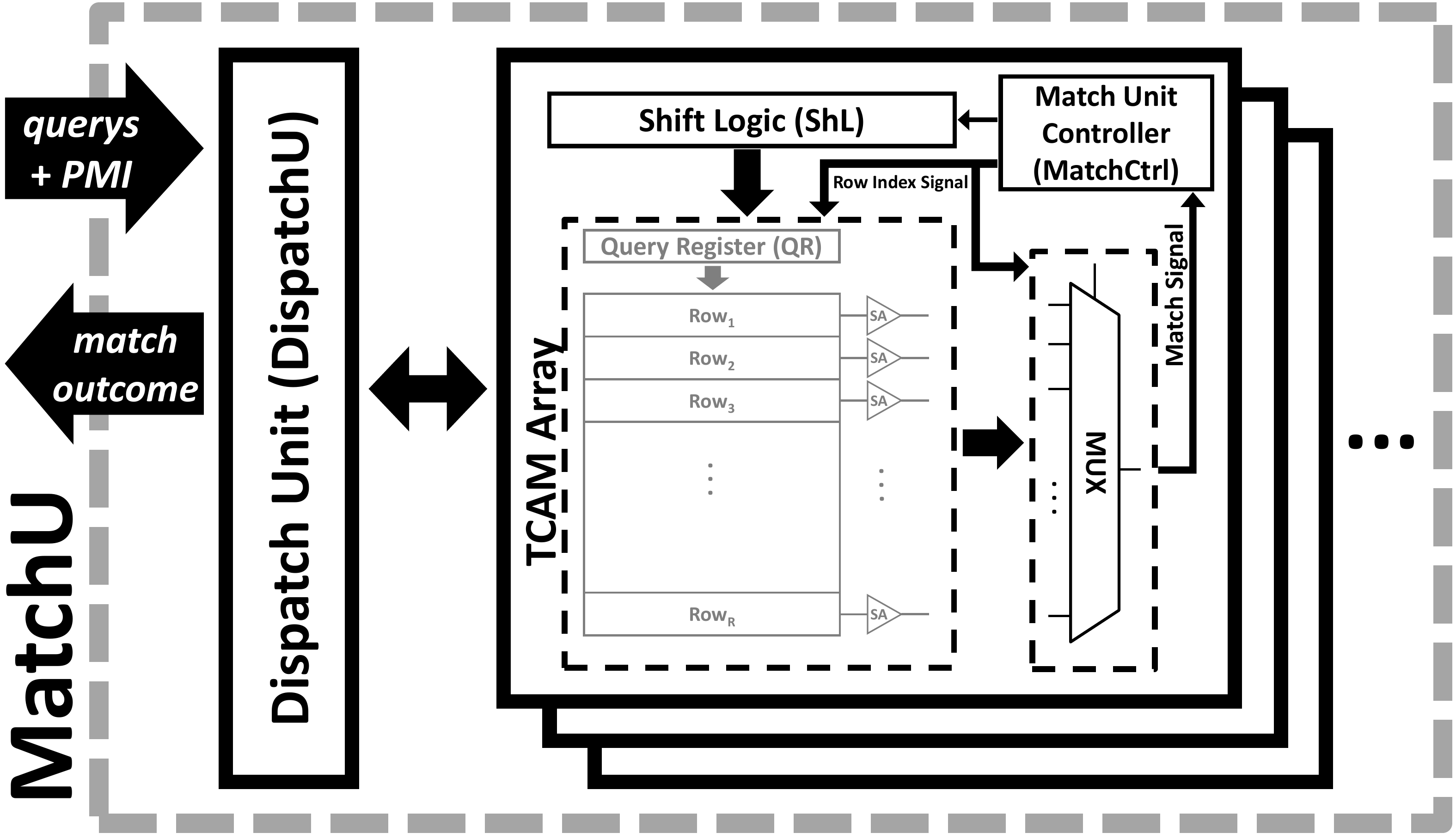}
\vshrink{0.1}
  \caption{Match Unit (MatchU)
  \label{fig:mu}}
\end{center}
\vshrink{.9}
\end{figure}

\begin{figure*}[htp]
  \begin{center}
  \includegraphics[width=.95\textwidth]{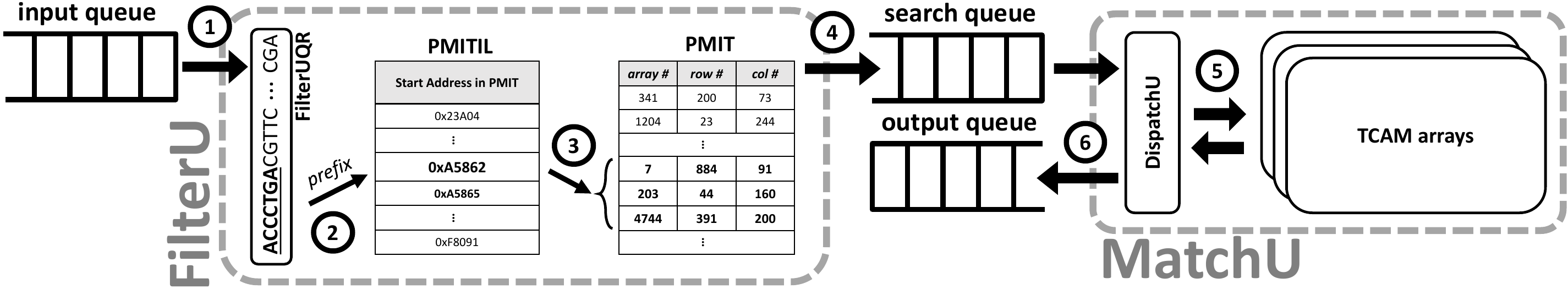}
  \vshrink{0.1}
  \caption{Life-cycle of a \ssq\ in \arch.
  \label{fig:example}}
\end{center}
\vshrink{0.8}
\end{figure*}

\subsubsection{Match Unit (MatchU)} 
\noindent Fig.~\ref{fig:mu} provides the structural organization of MatchU,
which orchestrates search. MatchU features the {\em Dispatch Unit} (DispatchU)
and non-volatile TCAM arrays capable of similarity search under NGS or genomic
variation induced noise.  DispatchU acts as a scheduler for TCAM search.
For each input \ssq\ to be mapped to the \srg, DispatchU collects the TCAM
$array_\#$, $col_\#$ and $row_\#$, as extracted
from the PMIT in FilterU, to initiate the targeted search. 

The input \ssq\ stays in the {\em Query Register} (QR) of the TCAM array
$array_\#$ during TCAM access. 
{\em Shift Logic} (ShL) in TCAM array $array_\#$ in turn first aligns the {\em prefix}
of length \seed\ of the \ssq\ with the {\seed}-long (matching) sub-sequence of
the \srg\ residing (in array $array_\#$) in row $row_\#$, starting from column
$col_\#$.  To this end, ShL shifts \ssq\ bits in QR and inserts Xs accordingly.
{\em Match Unit Controller} (MatchCtrl) orchestrates this operation.
Once alignment completes, MatchCtrl activates 
the row $row_\#$ for search.
Once the search completes, MatchCtrl provides
DispatchU with the indices of the \srg\ which demarcate the most similar
sub-sequence to the entire \ssq. {DispatchU then forwards these indices to the
output queue.}  

\begin{figure}[tp]
\vshrink{0.2}
\begin{center}
  \subfloat[{\em full match}]{  
	\includegraphics[width=0.42\columnwidth]{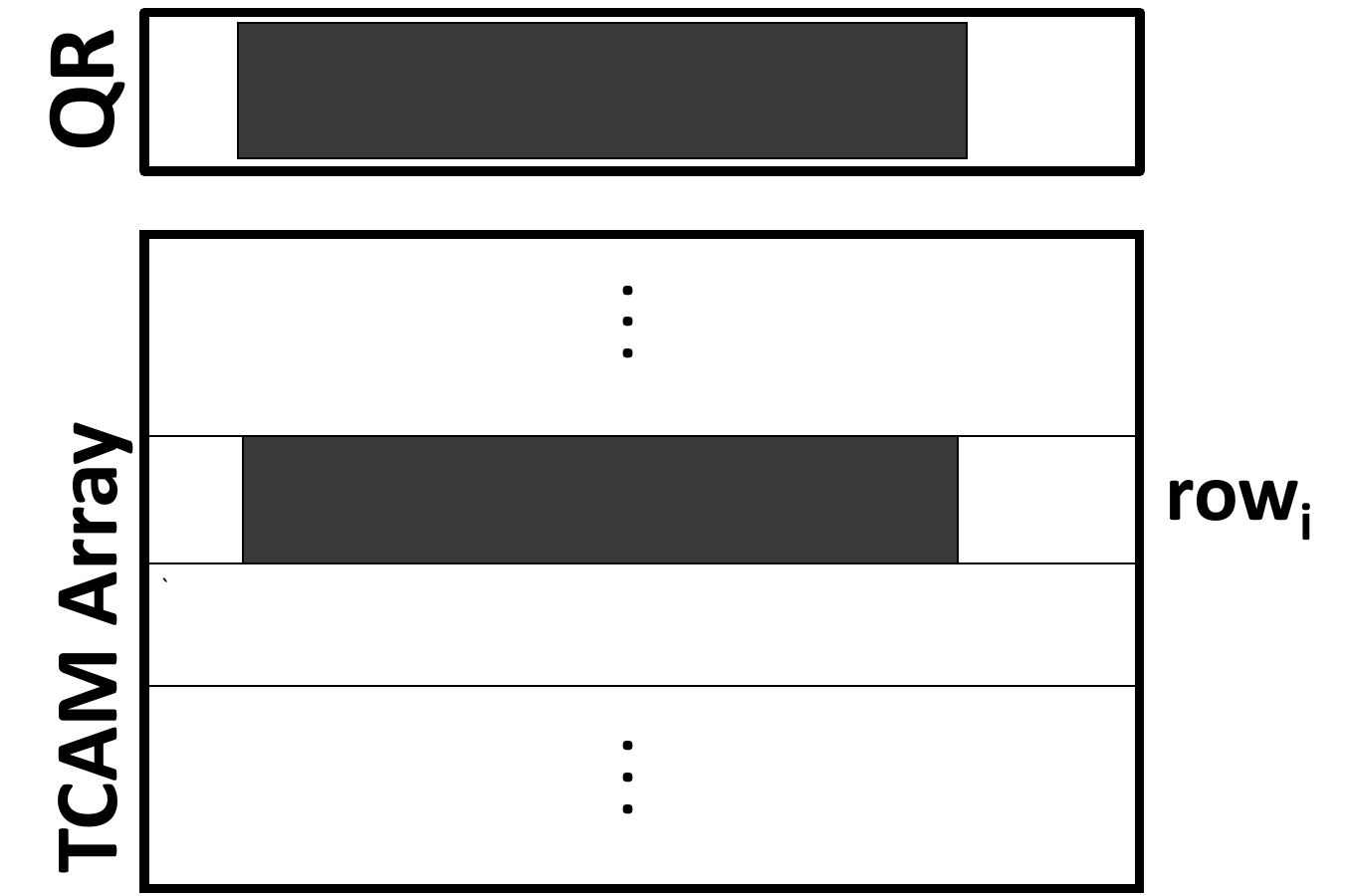}
		\label{fig:fullmatch}
 } 
  \subfloat[{\em fragmented match}]{  
	\includegraphics[width=0.42\columnwidth]{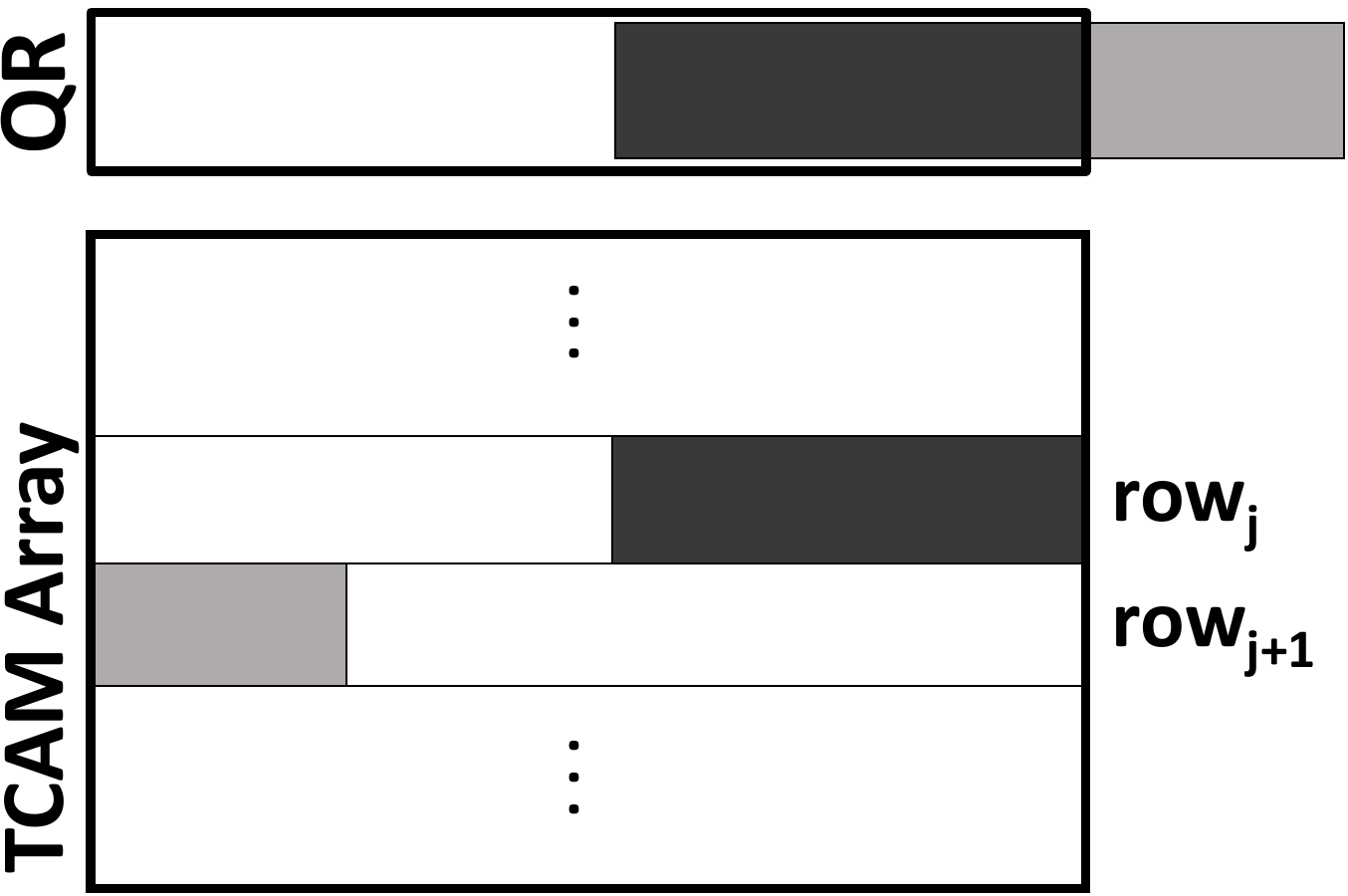}
		\label{fig:partmatchright}
  }
  \vshrink{0.1}
  \caption{Full (a) and fragmented (b) TCAM match.
  \label{fig:match}}
\end{center}
\vshrink{0.9}
\end{figure}

Fig.~\ref{fig:match} depicts two different match scenarios: In
Fig.~\ref{fig:fullmatch}, the \ssq\ (shown in dark shade within QR, white space
corresponding to Xs for padding) matches a sub-sequence of the \srg\ which is
entirely stored in a single row of the array. We call this scenario a {\em full
match}.  On the other hand, in Fig.~\ref{fig:partmatchright}, the \ssq\ matches
a sub-sequence of the \srg\ which is stored in two consecutive rows of the
array.  We call this scenario a {\em fragmented match}.  Fragmentation can
happen at both ends of the \ssq.  For example, in Fig.~\ref{fig:partmatchright},
the first portion of the \ssq\ (shown in darker shade) matches the end of row j,
while the rest (shown in lighter shade) matches the beginning of the next row,
row j+1.
MatchCtrl needs to address such fragmentation as \arch\  lays out the character
string representing the \srg\ in two dimensions in each array {\em
consecutively}. 

Conventional TCAM can only detect full match.  Handling fragmented match
requires extra logic.  By default, the TCAM array would select the longest
sub-sequence {\em l} of the \srg\ matching the input \ssq\ if a full match is
not the case, where {\em l} occupies an entire row. 
The darker-shade region in Fig.~\ref{fig:partmatchright} 
corresponds to such {\em l}. As {\em l} may be aligned to either the beginning
(Fig.~\ref{fig:partmatchright}) or the end 
of the \ssq, MatchCtrl has to additionally check the next or the previous row,
respectively, for a match to the unmatched portion of the \ssq.  We call the
first case a 
{\em fragmented tail  match}; the second, a {\em fragmented head match}.
In case of a fragmented match, search in the TCAM array takes two steps.
As a fragmented match may also happen at TCAM array boundaries, each array's
last row keeps the contents of the first row of the next array in sequence. 
\vshrink{0.1}


\subsubsection{Putting It All Together}
\label{sec:example}

\noindent 
Fig.~\ref{fig:example} summarizes the 6 steps in mapping a \ssq\ to the \srg:
First, FilterU retrieves a new \ssq\ from the head of the input queue at step
\raisebox{0.5pt}{\textcircled{{\raisebox{-0.9pt}{1}}}}.  In this case \seed =7
(bases) with the corresponding 7-base {\em prefix} of the \ssq\ underlined.
Then, at step
\raisebox{0.5pt}{\textcircled{{\raisebox{-0.9pt}{2}}}}, FilterU locates the
entry for the 7-base {\em prefix} of ACCCTGA in PMITIL, and extracts
the corresponding PMIT address(es).  Next,  at step
\raisebox{0.5pt}{\textcircled{{\raisebox{-0.9pt}{3}}}}, FilterU retrieves
TCAM array, column, and row numbers (i.e., $array_\#$,
$col_\#$, and $row_\#$; for targeted search in MatchU) for the
sub-sequences of the \srg\ which match the {\em prefix} ACCCTGA, from the PMIT addresses
collected at step \raisebox{0.5pt}{\textcircled{{\raisebox{-0.9pt}{2}}}}. 
Finally, FilterU sends the \ssq\ along
with $array_\#$, $col_\#$, and $row_\#$ to MatchU over the search queue at step
\raisebox{0.5pt}{\textcircled{{\raisebox{-0.9pt}{4}}}}.  At step
\raisebox{0.5pt}{\textcircled{{\raisebox{-0.9pt}{5}}}}, DispatchU initiates
search in TCAM array $array_\#$, at $row_\#$ and $col_\#$, and collects the
match outcome.  At step \raisebox{0.5pt}{\textcircled{{\raisebox{-0.9pt}{6}}}},
MatchU sends the match outcome to the output queue.


\ignore{
In this study, we only consider {\em substitution} errors, as other types of
errors are very rare in comparison to substitution errors~\cite{insdel} for
short {\ssq}s, similar in size to {\em read}s from modern Illumina NGS
platforms.

The most common form of mutation, {\em Single Nucleotide Polymorphism} manifests
itself as a substitution error, as well.
Therefore, {\em read mapping} by definition is after {\em similarity} rather
than an {\em exact match} between the input {\ssq}s and the \srg.  Accordingly,
for each input \ssq, \arch\ tries to locate the {\em most similar} sub-sequence
of the \srg\ to the \ssq, and returns the range of its indices. 
In this study, we only consider {\em substitution} errors, as other types of
errors are very rare in comparison to substitution errors~\cite{insdel} for
short {\ssq}s, similar in size to {\em read}s from modern Illumina NGS
platforms, which constitute 90\% of all reads in the world currently.  Insertion
and deletion errors become more prominent {for other sequencing platforms that
generate} significantly longer {\ssq}s.

For the proof-of-concept \arch\ design, we cap the {\em read} length by 100
bases and only consider substitution errors, as other types of errors are very
rare in comparison to substitution errors for such {\em short read}s (similar in
length to {\em read}s from modern Illumina NGS platforms, which constitute 90\%
of all reads in the world currently).  According to Illumina data-sheets,
machine-induced insertion and deletion errors (i.e., indels) in such short {\em
read}s of 100-200 bases are negligible.  However, {\em mutation-induced} indels
might still be present.  While such errors are still rare, it is important to
capture them, since this information is usually what we are after. 

In practice,  more than one {\em read} can contain, i.e., {\em cover} a given
base of the {\em reference} genome.  The average number of {\em read}s covering
any given base in the \srg\ is called {\em coverage} or {\em depth} of
sequencing. 
Mutation-induced variations (in the {\em read}s) from the {\em reference} genome
can take different forms such as substitution, insertion, deletion, duplication,
inversion, and interchromosal translocation. Many of these variations are very
difficult to detect in general. For example, to date, there is no
widely-accepted algorithm for detecting variations such as large indels,
duplications, or inversions.  Short (i.e., less than 50-base long) indel
variations represent the second frequent variations after mutation-induced
substitutions (i.e., SNPs). 
%
Although determining the exact frequency of short indels is not possible,
different studies estimate their frequency as approximately 1/10$^{th}$ of the
SNPs'~\cite{var1,var2,var3}.
Assuming a SNP frequency of 10$^{-3}$ (which represents a well-accepted number),
the  frequency of short indels would be 10$^{-4}$. In this case, for a {\em
read} length of 100 bases, each {\em read} would cover a short indel with
probability 10$^{-2}$. In other words, only approximately 10$^{-2}$ of the {\em
read}s would cover short indels.  
}


\section{\arch: Microscopic View}
\label{sec:pract}
\noindent 

\subsection{Search Space Pruning}
\label{sec:prune}
\noindent  In order to prune the search space, \arch\ first locates
sub-sequences of the \srg\ matching the \seed-long {\em prefix} of the \ssq\ in
FilterU (Section~\ref{sec:biocam}).  \seed\ represents a key \arch\ design
parameter which dictates not only the storage complexity, but also the degree of
search space pruning, which in turn determines \arch's throughput performance
and energy efficiency.

PMITIL grows with 4$^{seed}$, therefore, the larger the \seed, the higher
becomes the storage complexity. However, a larger \seed\ is more likely to
result in a lower number of {\em prefix} matches in the PMI tables, and hence, a
lower number of targeted searches in the MatchU.  In either case,  the \seed\
value
remains much less than the expected length of the \ssq. 

PMIT can have at most as many entries as the total number of {\em seed} long
sub-sequences 
contained {within the} \srg. This practically translates into the length of the
\srg, as a {\em prefix} can start from each base position of the \srg\ onward.
\ignore{
If PMIT is organized to keep multiple matches consecutively,
it suffices to keep per PMITIL entry just the start address in the PMIT for the
first match. The end address in this case simply corresponds to the start address of
the next PMITIL entry.
}
As PMIT is organized to keep multiple occurrences of the same {\em prefix} consecutively,
each PMITIL entry just keeps the start address in the PMIT for the
first occurrence.
PMIT, on the other hand, has to keep the $<$TCAM array
number, column number, row number$>$ tuple for each {\em prefix} match.  
If the
\srg\ is the human genome, PMIT would have approximately 3Giga entries. As we
will detail in Section~\ref{sec:setuptbl}, 32 bits suffice to store each $<$TCAM
array number, column number, row number$>$ tuple per PMIT entry; and 32 bits,
each $<$PMIT start address$>$ per PMITIL entry.

\ignore{
To improve throughput performance, after generating PMIT, \arch\ shuffles the
order of entries as follows:
\begin{list}{\labelitemi}{\leftmargin=1.0em} 
  \item} 
	
	PMIT keeps the entries corresponding to the very same {\em prefix}
	always at consecutive addresses, and re-orders such entries further to have
	all entries pointing to the same TCAM array reside at consecutive addresses.
	\arch\ processes multiple PMIT matches per {\em prefix}  in this consecutive
	order. 
Under such re-ordering, 
communicating a list of PMIs 
and performing search 
in the array
happen in a pipelined fashion. This masks
communication
latency and consequently, can improve throughput performance significantly.

\ignore{
\arch\ re-orders the PMIT entries to have search requests to different
  TCAM chips interleaved. In other words, \arch\ tries to avoid sending multiple
consecutive search requests to the very same TCAM chip to maximize (TCAM)
chip-level parallelism.
\end{list}
\vshrink{0.1}}

\subsection{Data Representation}
\label{sec:encode}
\noindent
Each input \ssq\ and the \srg\ itself represent character strings over the alphabet
\{A, G, C, T\}. Conventional bioinformatics formats such as FASTA~\cite{fasta}
encode each letter from such alphabets of bases
by single-letter ASCII codes. However, TCAM arrays conduct the search at bit
granularity. Therefore, \arch\ needs to translate {\em base character} mismatches to {\em
bit} mismatches.
To this end, \arch\ adopts an encoding which renders the very same number of
mismatched bits for a mismatch between any two base characters.  
This would not be the case, if we encoded each base character in \{A, G, C, T\}
by simply using 2 bits: a base-mismatch would sometimes cause a 2-bit mismatch
(e.g., when comparing `01' to `10');
other times, a single-bit mismatch (e.g., when comparing `00' to `10'). 
\arch's encoding instead uses 3 bits per base character, where
any two 3-bit code-words differ by exactly 2 bits, 
such as \{111, 100, 010, 001\}. Thereby \arch\ guarantees that exactly 2
bits would mismatch for any base character mismatch.

\subsection{Similarity Search}
\label{sec:similar}
\begin{figure}[tp]
\vshrink{0.1}
  \begin{center}
  \includegraphics[width=0.3\textwidth]{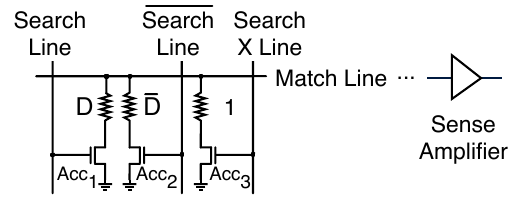}
  \vshrink{0.2}
  \caption{Resistive TCAM cell~\cite{ipekTCAM}.
  \label{fig:tcamCell}}
\end{center}
\vshrink{0.8}
\end{figure}

\noindent Fig.~\ref{fig:tcamCell} depicts a representative resistive TCAM cell.  The two
resistors attached to the access transistors $Acc_1$ and $Acc_2$, respectively,
carry the data bit value $D$ and its complement $\overline{D}$.  The high (low)
resistance value $R_{high}$ ($R_{low}$) encodes logic 1 (0).  The third resistor
attached to $Acc_3$ is hardwired to logic 1, hence its resistance remains
constant at $R_{high}$. 

To search for logic 0, {\em Search Line} ($SL$) is set to 0, and its complement
$\overline{SL}$ to 1, such that $Acc_2$ turns on; $Acc_1$ off.  Thereby only the
resistor carrying $\overline{D}$, $R$, gets connected to the {\em Match Line}
($ML$).
If the cell content was 0, i.e., ${D}=0$ and $\overline{D}=1$, there would be a
match, and $R=R_{high}$ applies.  Otherwise, if the cell content was 1, i.e.,
${D}=1$ and $\overline{D}=0$, there would be a mismatch, and $R=R_{low}$
applies.  A symmetric discussion holds for searching for logic 1. 
On a per TCAM cell basis, $R_{high}$ connected to $ML$ indicates a match,
$R_{low}$, a mismatch.
To search for X, both $SL$ and $\overline{SL}$ are set to 0, and {\em Search X
Line} 
to 1,  such that only the hard-wired $R_{high}$ attached to $Acc_3$ is connected
to $ML$.  This is how search for X always renders a match, independent of the
value of $D$.
Each cell within each row contributes to the effective resistance connected to
$ML$, $R_{eff}$,  by  $R_{high}$ ($R_{low}$) on a match (mismatch).  The {\em
Sense Amplifier} SA (connected to the $ML$) in each row signals a (mis)match for
the entire row depending on the value of $R_{eff}$.  SA would only signal a
row-wide match, if all cells match, i.e., if each cell contributes to $R_{eff}$  by
$R_{high}$. Let us call the $R_{eff}$ in this case $R_{row-wide-match}$.  SA would signal
a row-wide mismatch if at least one cell mismatches, i.e., contributes to $R_{eff}$ by
$R_{low}$. The value of $R_{eff}$ in this case evolves with the number of
cell mismatches, and assumes the closest value to $R_{row-wide-match}$ under a single-cell
(bit) mismatch. 
The contents of the TCAM query register directly correspond to the values on
the search lines.  

In a TCAM array based on the cell from Fig.~\ref{fig:tcamCell}, unless all bits
within a row match, SA always signals a mismatch for the entire row.
However, as explained in Sect.~\ref{sec:pr},
a matching {\em query} may indeed have a few bases
different from the corresponding sub-sequence of the {\em reference}, due to NGS
or genomic variation induced noise.  To
resolve this discrepancy, \arch\ deploys tunable SAs which associate a wider
$R_{eff}$ range with a row-wide match. 
We can tune these SAs to signal a row-wide match when less than a given number
{\em t} of bits mismatch, which translates into less than {\em t} $R_{low}$s
connected to $ML$. 
We will refer to {\em t} as the {\em tolerance}, which represents
an adjustable design parameter. 

The gap between $R_{eff}$ levels corresponding to different number of
mismatching bits decreases as the number of mismatching bits grows, complicating
SA design.  At the same time, due to PVT (Process, Voltage, Temperature)
variations, individual TCAM cell resistance levels may notably deviate from
nominal $R_{high}$ or $R_{low}$, leading to divergence of such $R_{eff}$ levels
from their expected values.  In Sect.~\ref{sec:sensest}, we will detail how
\arch\ tunes the SAs in a variation-aware fashion.

Using \arch's 3-bit (per base) encoding  (Sect.~\ref{sec:encode}), any single-base mismatch
results in exactly 2-bit mismatches. Under the more intuitive 2-bit encoding, on the
other hand, a base-mismatch
sometimes causes 2-bit mismatches; other times, a single bit mismatch,
further complicating the configuration of {\em tolerance}.



\subsection{Hierarchical Multi-Phase Search}
\label{sec:hier}
\noindent
Our focus so far was on the very basics of \arch's mapping mechanism.  We will
next look into the mapping accuracy, specifically, under what circumstances
\arch\ may not be able to map a given {\em query} to the respective {\em
reference}, which in fact was {\em similar enough}. In the following, we will
refer to such cases as {\em missed} mappings. 
NGS imperfections (i.e., {read errors}) and genomic variations complicate
mapping, and thereby can lead to misses. 

\begin{figure}[tp]
\begin{center}
  \includegraphics[width=0.37\textwidth]{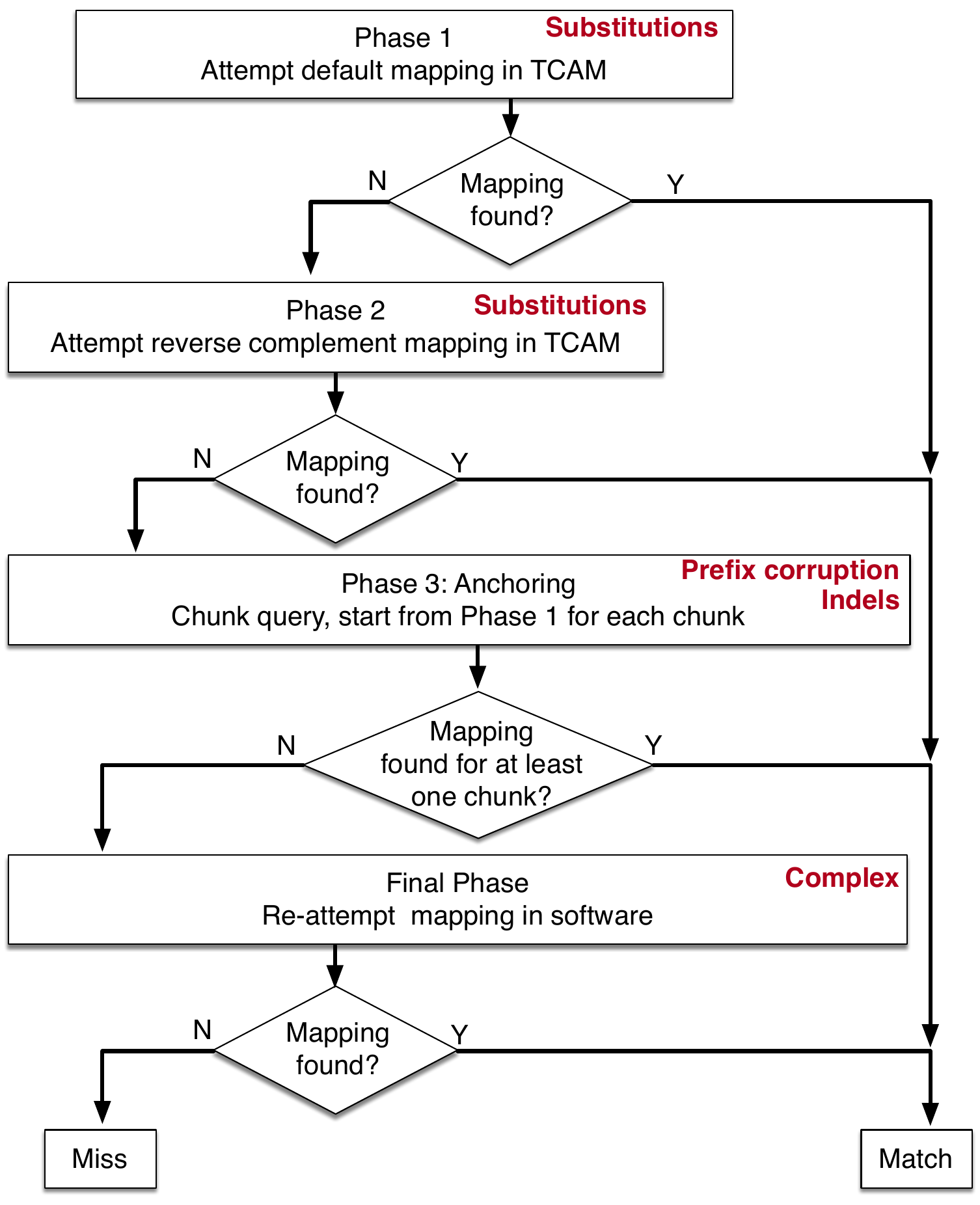}
\vshrink{0.1}
  \caption{\arch's hierarchical multi-phase search flow.
  \label{fig:phase}}
\end{center}
\vshrink{0.9}
\end{figure}

As explained in Sect.~\ref{sec:pr}, \arch\ is designed to operate under all
prevalent manifestations of {read errors} and genomic variations. To this end,
\arch\ employs multi-phase hierarchical mapping. Fig.~\ref{fig:phase} provides
an overview. Each phase acts as a filtering layer for the subsequent phase,
which in turn performs more complex mapping.
{\em More complex mapping} entails re-attempting by considering {\em more
complex manifestations} of {read} errors/genomic variations than the predecessor
phase did.  In this manner, each phase re-attempts mapping only for the subset
of {\em query}s that the previous phase missed to map.  

If the mapping in a phase fails, MatchU raises the {\em Missed-Map} signal.
\arch\ in turn feeds {\em Missed-Map} back to FilterU, to trigger  more
complex mapping attempts in the subsequent phase(s). 

\vshrink{-0.1}
\noindent{\bf Phase 1} attempts a
basic mapping in the TCAM arrays, assuming that a reverse
complement (Sect.~\ref{sec:pr}) is not the case.  In Phase 1, FilterU and MatchU
closely follow the steps detailed in Sect.~\ref{sec:biocam}.
\arch\ TCAM search, by construction, can effectively
identify similarity under base substitutions, which represent the common case
per Fig.~\ref{fig:pie}. %
Phase 1 can miss a mapping under three cases:
\vshrink{.1}
\begin{list}{\labelitemi}{\leftmargin=1em}
  \itemsep-0.2em
  \item[\bf (i)] {\bf Reverse complement:} the {\em query} is a {\em read} sequenced in reverse order. The
	probability for this case, $P(i)$, is approximately 50\%.
  \item[\bf (ii)] {\bf Prefix corruption:} the {\em query}'s \seed-long {\em prefix} (used for search space pruning
  in FilterU) has substitutions or indels. A corrupted {\em
prefix} may lead to ill-addressed search requests, i.e., FilterU sending
incorrect PMIs to MatchU. 
If the probability of
having a corruption in a given base location is $P(loc)$, 
the probability for this case, $P(ii)$, becomes 
$1-(1-P(loc))^{seed}$. 
We can estimate $P(loc)$ by adding a typical {read}
error rate of 0.1\%~\cite{insdel} to an average genome variation rate of 0.1\%~\cite{var1,var2,var3}.
Using this  estimate, for a representative \seed\ value of 15 (Sect.~\ref{sec:setup}),
$P(ii)$ barely reaches 3.0\%.
  \item[\bf (iii)] {\bf Indels:} the {\em query} contains indels, anywhere. The most common
  indels are short indels induced by genome variations.
Let $P(indel)$ be the probability of a short indel, and $len$, the length
of the {\em
query}. Then,
$P(iii) = 1-(1-P(indel))^{len}$
applies. 
While there is no consensus on $P(indel)$,
$0.01\%$ represents a conservative estimate~\cite{var1,var5}, which  
renders
$P(iii) \approx 1.5\%$ 
for a typical {\em query} length of 150~\cite{illumina}. 
\end{list}
\vshrink{0.1}


\noindent{\bf Phase 2} handles missed mappings due to reverse complements.
After getting {\em Missed-Map} from MatchU (at the end of Phase 1), FilterU
immediately sends PMIs corresponding to the reverse complement of the {\em
query} to MatchU. To accelerate processing, MatchU employs an extra register
inside the Shift Logic, which keeps the reverse complement of the {\em query} in
addition to the original.  MatchU copies the reverse complement in this register
at the time it gets 
the original {\em query} (during Phase 1).  Therefore, upon receipt of {\em
Missed-Map}, FilterU does not need to broadcast the reverse complement
separately, but only the PMIs for the reverse complement (which FilterU simply
extracts by consulting the PMI tables with the \seed-long {\em prefix} of the
reverse complement.)

\vshrink{-0.1}
\noindent{\bf Phase 3} handles missed mappings due to {\em prefix} corruptions
and indels,
by adapting 
{\em anchoring}~\cite{bowtie1}.  This phase processes all {\em query}s
which Phase 2 was not able to map. 
Phase 3 first anchors the {\em query} in the middle to chunk the {\em query}
uniformly into two. Then, each chunk separately goes through Phase 1, and if
necessary, through Phase 2. 
Unless Phase 1 (or Phase 2, as need be) manages to map at least one of the two
chunks 
to the {\em reference}, Phase 3 is considered to miss
the mapping.

\begin{figure}[tp]
\begin{center}
  \includegraphics[width=\columnwidth]{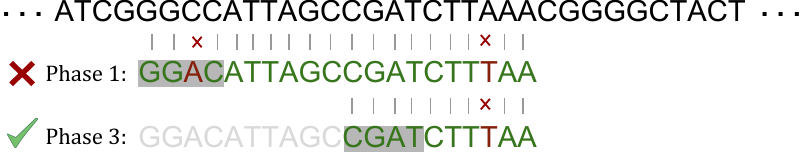}
\vshrink{0.5}
  \caption{An example of anchoring.
  \label{fig:anchoring}}
\end{center}
\vshrink{0.8}
\end{figure}

Fig.~\ref{fig:anchoring} depicts an example where 
Phase 3 maps a {\em read} which Phase 1 
fails to map due to 
prefix corruption.  The top row depicts the relevant portion of the {\em
reference}.
The second and third rows show the corresponding {\em read}, with pointers to
the matching outcome at Phase 1 and 3, respectively.  The alignment of the
{\em read} w.r.t. the {\em reference} reflects the ideal alignment (which
renders the most similar mapping). The shaded
portion corresponds to the {\em prefix} (of length 4 in this case). The {\em
read} and {\em prefix} lengths are not representative, but ease
illustration.
Phase 1 fails to identify this alignment due to the single base corruption
(C$\rightarrow$A) in the  third base of the {\em prefix}. Phase 2 is of not much
help either, as a reverse complement is not the case. 
Phase 3 comes to rescue, by chunking, i.e., {\em anchoring} the {\em read} in the middle, and
attempting mapping for each half. The mapping of the first half still fails in
this case, due to the very same corruption in the {\em prefix}. The {\em prefix}
of the second half (i.e., CGAT), however, is not corrupted, and \arch\ TCAM arrays can easily
handle the single base mismatch in this half, which renders the correct alignment as a result
-- simply following the alignment dictated by the second half for the entire
{\em read}.

\ignore{
In this case, Phase 3 comes to rescue, by trying to map the 
second half of the read. As depicted, this time \arch\ successfully maps 
the read, as prefix of the second half is not corrupted and TCAM arrays 
can handle a single base mismatch.
}

Similar to this example, the proof-of-concept \arch\ design adapts 2-way
chunking for anchoring, where multi-way (and not necessarily uniform) chunking
can further help reduce the number of missed mappings. 
Anchoring essentially is a {\em divide and conquer} method. The key insight is
that corruptions to lead to missed mappings are much less likely to occur in all
of the shorter chunks simultaneously.
In Sect.~\ref{sec:acc}, we will demonstrate how anchoring can improve mapping
accuracy significantly. 

Phase 3 can miss a mapping under the very same two conditions as Phase 1 (namely
{\em prefix} corruptions and indels anywhere; cases {\bf (ii)} and {\bf (iii)}),
but only if these conditions apply to both of the chunks under {\em anchoring}.
Under uniform two-way chunking, a typical {\em query} length of 150 bases
renders a chunk length of 75, for which $P(ii)$ and $P(iii)$ become 3.0\% and
1.0\%, respectively.  
Hence, the probability to miss the mapping of a chunk would be approximately
4.0\%. As Phase 3 can miss a mapping only by missing both chunks, the
probability of missing a mapping in Phase 3 becomes approximately $(4.0\%)^2 =
0.16\%$.

By construction, anchoring can also help with more complex scenarios than
indels or prefix corruptions, including long-indels, gaps, base duplications or
inversions, as long as one half of the affected querys is still mappable 
by \arch. 
%
While anchoring reduces the miss probability significantly,
Phase 3 may still miss mappings due to rare complex variations.
To map 
such problematic
{\em query}s, \arch\ can always be paired with
sophisticated software algorithms.

\ignore{

\footnote{
\noindent \arch\ also relies on post-processing by software
to re-format the match outcome
for mapped {\em query}s, i.e., to find different SAM format fields such as CIGAR
and MAPQ.
}. 

\arch\ stores such missed {\em query}s 
in a dedicated memory location
for the software to process.
We will quantify the
overhead of this hardware-software hybrid solution, \arch++, 

in Sect.~\ref{sec:eval}
}

\ignore{
\subsection{System Integration}
\label{sec:si}
\noindent 
{
  Without loss of generality, all components of the proof-of-concept \arch\
implementation reside in a single card attached to the PCIe bus.
The host loads the {\em reference}, PMIT, and PMITIL tables to the \arch\ card,
before
starting mapping by sending \arch\ the range of addresses of {\em read}s inside
the main memory, read length, and a dedicated memory space to write back the results.}
%

While \arch\ features non-volatile TCAM data arrays as both of ~\cite{ipekTCAM, ipekACC} do,
\arch\ arrays do not include any of the priority index logic, population count
logic, or the reduction network from Guo et al.~\cite{ipekTCAM} or programmable
microcontrollers of ACC-DIMM~\cite{ipekACC}.
Instead, \arch\ tailors the data and control paths to {\em read mapping}, which
entails minimal logic for filtering, scheduling, and queuing search requests.
At the same time, \arch's TCAM arrays incorporate a novel similarity match
mechanism cut for {\em read mapping}.
%

}

\section{Evaluation Setup}
\label{sec:setup}

\subsection{{System-level Characterization}}
\label{sec:systemSetup}
\noindent Without loss of generality, all components of the proof-of-concept \arch\
design reside in a single card attached to the PCIe bus.
The host loads the {\em reference}, PMIT, and PMITIL tables to the \arch\ card
before mapping starts. 
\ignore{
starting mapping by sending \arch\ the range of addresses of {\em read}s inside
the main memory, read length, and a dedicated memory space to write back the results.
}

We evaluate \arch\ using $N$ different FilterUs and MatchUs to increase
parallelism, as shown in Fig.~\ref{fig:setupOrg}. 
Each MatchU stores $1/N$ of the reference genome.
Therefore, having multiple MatchUs does not need 
extra TCAM space, but needs more NOC resources to feed all MatchUs in parallel. 
On the other hand, each FilterU is dedicated to a single MatchU, covering 
potential matching indices corresponding to the part of the genome stored in that MatchU.
Hence, we divide PMIT table, as well, into smaller banks each corresponding to 
a FilterU, and a MatchU consequently.
Having multiple PMIT banks for each FilterU only needs chunking, and no extra space.
Besides, we need a separate PMITIL table for each PMIT bank. As discussed in 
Section~\ref{sec:biocamorg}, size of PMITIL tables depends on the {\em seed} only,
not the size of PMIT. Then, having $N$ FilterUs and MatchUs increases the space
needed to store PMITILs by a factor of $N$.

\begin{figure}[tp]
  \begin{center}
  \includegraphics[width=\columnwidth]{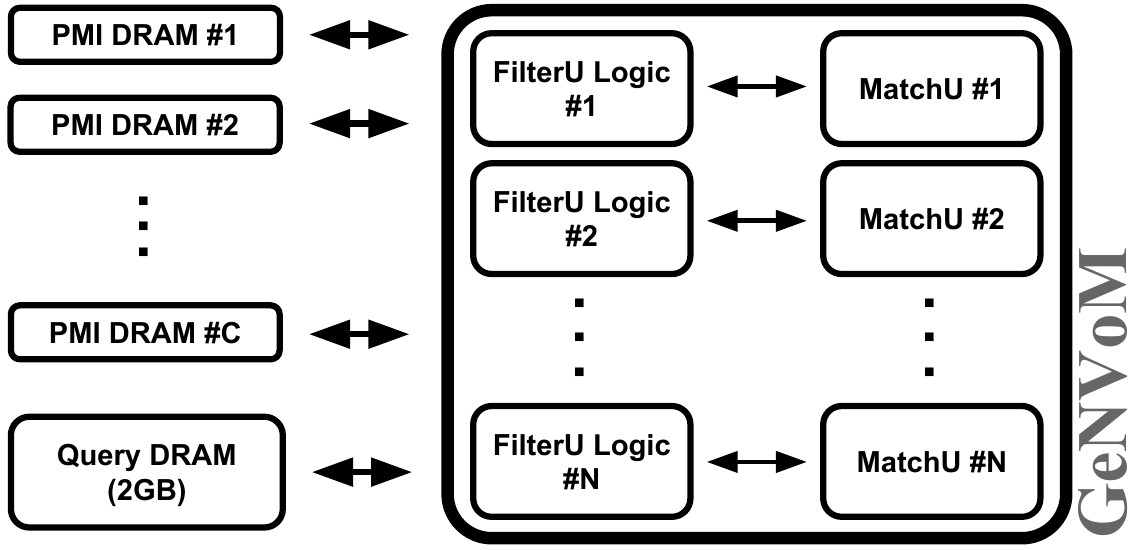}
  \caption{Organization of a single \arch card.
  \label{fig:setupOrg}}
\end{center}
\vshrink{1}
\end{figure}

A total on-card DRAM space of size $M$ Giga bytes stores all PMIT and PMITIL tables.
The $M$ Giga bytes of memory is connected to \arch\ via $C$ channels, 
to increase bandwidth of loading PMIs to meet the needs of FilterUs. 
A dedicated memory of 2 Giga bytes is used to store the querys.
Each \arch\ design is characterized by the four 
parameters, $M$, $C$, $L$, and $N$ in the evaluation section, where $L$ represents
the {\em seed} value.
We sweep these 4 parameters and quantify throughput and energy-efficiency.

\arch\ relies on a GPU kernel (based on \cite{gpusw})
to re-format the match outcome of mapped {\em query}s, i.e., 
to find different SAM format fields such as CIGAR and MAPQ.
We run the kernel on an NVIDIA K40 node in parallel, where it
processes mapped {\em query}s in a pipe-lined manner periodically. 
This significantly reduces throughput
overhead of running the GPU kernel, while we quantify
its energy-efficiency implications by real measurements (similar 
to Section~\ref{sec:base}).

We use a modified version of Ramulator~\cite{ramulator} for simulation.  
While the default Ramulator can model accesses to the DRAM chips, 
we implement intra-card interactions during search, control logic and network related operations. 
We model LPDDR4-4266 DRAM memory to store PMIT, PMITIL tables, and the querys.
We use DRAMPower~\cite{drampower}
to estimate DRAM power consumption. DRAMPower does not simulate LPDDR4. Therefore,
we conservatively estimate DRAM power using DRAMPower's LPDDR3 model.

\ignore{
We evaluate \arch\ using 3 different TCAM chips, all containing 1K$\times$1Kbit arrays.
These chips have different capacities;
512Mbit, 1Gbit, and 2Gbit, respectively.  
We use a human genome as
the \srg\ (Section~\ref{sec:genome}), which has approximately 3Giga-bases.  As
discussed in Section~\ref{sec:encode}, \arch\ adopts a 3-bit representation for
each base.  
Therefore, a total of 16, 8, and 4 TCAM chips
find place on a single \arch\
DIMM, to store the 9Gbit \srg, 
considering the 3 different chip configurations respectively.  
We implement PMI tables as DRAM modules, and
keep them in the main memory in a separate DIMM such that the host can claim
DRAM space back as part of the main memory, as need be. All \arch\ logic and
controllers (from FilterU and MatchU) reside in the corresponding DIMM
controllers.  Another option (not considered in the evaluation) is packing PMI
tables into the same DIMM as the TCAM chips, to have a self-contained \arch\
DIMM (of possibly higher energy efficiency and throughput by optimizing
intra-DIMM communication).  {We do not practice this option to avoid favoring
\arch\ over the baseline for comparison (Section~\ref{sec:base}).} We use a
modified version of DRAMSim2~\cite{dramsim2} for simulation.  While the default
DRAMSim2 can model inter-DIMM communications, we had to implement intra-DIMM
interactions during search, control logic and network related operations. 
We model a DDR4 DRAM to store PMIT and PMITIL tables.

A larger {\em seed} decreases the number of false positives (which improves throughput), 
and at the same time, requires more memory space exponentially. Hence, in our experiments,
we pick the largest possible {\em seed} considering the memory space available to FilterUs.

}

\subsection{PMI Table Generation}
\label{sec:setuptbl}
\noindent As explained in Section~\ref{sec:prune}, PMIT keeps an entry 
for each possible \seed-long {\em prefix} contained within the \srg.
In other words, PMIT keeps an entry for each possible base position in the \srg\
to demarcate the start of a \seed-long {\em prefix}.
Therefore, PMIT capacity 
becomes practically independent of the {\em seed} for feasible {\em seed}
values.  Allocating an entry for each possible base position in the \srg, a
tight-enough upper bound for PMIT capacity for the human genome used as the
\srg\ for evaluation (Section~\ref{sec:genome}) is approximately 11.4GB,
independent of the \seed. As explained in Section~\ref{sec:systemSetup},
having $N$ PMI banks does not need extra space.

On the other hand, PMITIL contains 4$^{seed}$
entries. We evaluate \arch\ considering different \seed\ values. 
Table~\ref{tbl:pmiil} captures PMITIL capacity for practical \seed\ 
values ranging from 10 to 15.  
PMIT size and $N\times$PMITIL size together, $M$, determine the DRAM space
requirement of the proof-of-concept \arch\ implementation.

\begin{table}[tp]
  \resizebox{\linewidth}{!}{
\centering
\begin{tabular}{c|cccccc|}
\cline{2-7}
                                    & \multicolumn{6}{c|}{\textbf{Size (GB)}}                                           \\ \hline
\multicolumn{1}{|c|}{\textbf{\em seed}} & \textbf{10} & \textbf{11} & \textbf{12} & \textbf{13} & \textbf{14} & \textbf{15} \\ \hline
\multicolumn{1}{|c|}{PMITIL}         & 0.004       & 0.017       & 0.067       & 0.268       & 1.074       & 4.295     \\ \hline
\end{tabular}
}
\caption{{PMITIL table capacity as a function of \seed.} \label{tbl:pmiil}}
\vshrink{0.6}
\end{table}

\subsection{Circuit-level Characterization}
\label{sec:circuitSetup}
\noindent  
The proof-of-concept \arch\ implementation uses Phase Change Memory (PCM) as the
resistive memory technology for TCAM arrays, which features a relatively high
$R_{high}$ to $R_{low}$ ratio: 11.5 on average~\cite{ibmpcm2015}.  A higher
$R_{high}$ to $R_{low}$ ratio eases sensing (i.e., distinguishing between
matches and mismatches as explained in Section~\ref{sec:similar}), therefore,
enables arrays with longer rows.  \arch's TCAM arrays are similar to the most
energy-efficient design from Guo et al.~\cite{ipekTCAM}, which corresponds to a
1K$\times$1Kbit configuration.

We synthesize logic circuits by Synopsys Design Compiler vH2013.12 using the
FreePDK45 library~\cite{pdk45}. To match the technology of our baselines
for comparison (Section~\ref{sec:base}), we scale the outcome from 45nm to 28nm
using ITRS projections~\cite{itrs}.  
\arch's logic operates at 1GHz.
A single search operation takes 1ns to
complete, while consuming 0.1nJ of energy. For TCAM array area estimates,
we scale numbers from ~\cite{ipekTCAM} to 28nm.
We use ORION2.0~\cite{orion} to model the network.  Intra-DIMM H-tree network
operates at 1GHz, while each hop (1 router + link) consumes 3.83mW.
 



\subsection{Similarity Matching Specification}
\label{sec:sensest}
\noindent \arch\ adopts the Voltage Latch Sense Amplifier (VLSA) design
from~\cite{senseamp} to implement tunable sensing as explained in
Section~\ref{sec:similar}. We simulate VLSA in HSPICE v2015.06 using the
FreePDK45~\cite{pdk45} library.  VLSA's threshold voltage sets the boundary
between the ranges of the effective resistance, $R_{eff}$, values the SA
perceives as a (row-wide) match or a mismatch. We configure VLSA's threshold
voltage to account for potential fluctuation in $R_{high}$ and $R_{low}$ values
due to PVT variations.

\subsubsection{Setting the sensing threshold} 
\label{sec:th} 
\noindent We conduct a Monte Carlo analysis using the (variation-afflicted) $R_{high}$ and
$R_{low}$ distributions from IBM~\cite{ibmpcm2015},  extracted from measured data:
$\mu(R_{high})=243.8K\Omega$, $\sigma(R_{high})=50.9K\Omega$,
$\mu(R_{low})=21.2K\Omega$, and $\sigma(R_{low})=2.5K\Omega$.  $\mu$ and
$\sigma$ represent the mean and the standard deviation.  Considering a row size
of 1Kbits, we find $R_{eff}$ for 1M sample scenarios each corresponding to a
different number of base mismatches.  Using the resulting $R_{eff}$
distribution, and capping the maximum number of base mismatches that are
permitted to pass as a match (i.e., the {\em tolerance} as explained in
Section~\ref{sec:similar}), we set SA's sensing threshold in a variation-aware
manner, in a way to make sure that SA will not produce false negatives. 
In other words, we make sure that SA will never signal a mismatch, when query
is similar-enough.

\ignore{
\begin{table}[tp]
\resizebox{\linewidth}{!}{
\centering
\begin{tabular}{c|ccccc|}
\cline{2-6}                                                        & \multicolumn{5}{c|}{\textbf{ratio of matches missed (\%)}}     \\ \hline
\multicolumn{1}{|c|}{\textbf{\em tolerance}} & \textbf{0} & \textbf{1} & \textbf{2} & \textbf{3} & \textbf{4} \\ \hline
  \multicolumn{1}{|l|}{Genome Analyzer (error rate = 1\%)}               & 63.4       & 26.4       & 7.9        & 1.8        & 0.3        \\
\multicolumn{1}{|l|}{HiSeq 2500 (error rate = 0.1\%)}          & 9.5        & 0.5        & 0.02       & 0.0        & 0.0        \\ \hline
\end{tabular}
}
\caption{Ratio of actual matches missed due to NGS errors, considering two
  different, representative platforms. The maximum number of base mismatches
  that are permitted to pass as a match, i.e., the {\em tolerance}, is a
  critical 
  \arch\ parameter.\label{tbl:sa}
}
\vshrink{0.3}
\end{table}

\subsubsection{Setting the {tolerance}} 
\label{sec:tol} 
\noindent {\em Tolerance} is a critical design parameter for \arch.  Modern NGS platforms
fall short of sequencing DNA fragments with 100\% accuracy.  Therefore,  {\em
tolerance} should be set as a function of the expected 
(output) error rate of the target NGS platform.  As an example, let us assume a
NGS read error rate of 1\%, which corresponds to the probability of error in
each base position of any \ssq\ sequenced (Section~\ref{sec:pr}).  By
construction, \arch\ would {\em miss} (i.e., would not be able to map to the
\srg) any \ssq\ with too many read errors, particularly if the {\em tolerance}
remains too low. For an error rate of 1\%, the probability of having more than 1
base of a 100-base-long \ssq\ corrupted, just due to read errors, becomes:

\vshrink{0.3} $$1-\sum_{i=0}^{1} {100 \choose
i}(0.01)^i(0.99)^{(100-i)}=26.4\%$$ \vshrink{0.2}

Let us assume that the {\em tolerance} is set to 1 (base mismatch).  This means,
\arch\ would not map any {\ssq} having more than 1 base mismatch (with the
\srg).  Following the previous example, the input {\ssq}s to \arch\ would already have
more than 1 base corrupted with a probability of 26.4\%, which \arch\ would
capture as the equivalent of ``more than 1 base mismatch''.  In other words,
with a probability of 26.4\%, \arch\ would miss a \ssq\ for mapping, which would
have been successfully mapped to the \srg\ were there no read errors.
Table~\ref{tbl:sa} gives the ratio of the actual matches missed in this manner,
for a \ssq\ length of 100 bases, considering the read error rates of two
  representative NGS platforms from Illumina~\cite{illumina}, HiSeq 2500 and
  Genome Analyzer, respectively.  Similar tables can help in setting the {\em
  tolerance}, which in turn determines the threshold for SA.
}

\subsubsection{Sensing Accuracy}
\label{sec:fp} 
\noindent As explained in Section~\ref{sec:similar}, under PVT variations, sense
amplifiers may trigger 
a (row-wide) match in case of an actual (row-wide) mismatch.  For each 
such case, the number of base mismatches remains higher than the preset {\em
tolerance} value. In the following, 
we will refer to
this difference in the number of base mismatches with respect to the {\em
tolerance} as {\em overshoot}.  
These cases are not necessarily errors, and rather
translate
into a \ssq\ of less similarity than expected being matched to a sub-sequence
of the \srg.
Therefore, as long as the overshoot (in terms of base mismatches) with
respect to the anticipated  {\em tolerance} remains bounded, each such case
can easily pass as a {\em less similar match} (which in fact can be an actual match
where the input {\em query} was significantly corrupted). 
Monte Carlo analysis from Section~\ref{sec:th} shows that for different
representative {\em tolerance} values used in Section~\ref{sec:eval}, 
overshoot is usually less than 3, with
probability of an overshoot of size 3 or larger barely reaching 0.05\%. 
{
We quantify the impact of SA's inaccuracy on \arch's overall accuracy 
in Section~\ref{sec:acc}.
}

%

\ignore{
Fig.~\ref{fig:sa} quantifies the probability of such {\em less similar match} cases,
along with the overshoot in the number of base mismatches with
respect to the {\em tolerance}, for the two NGS read error rates from
Table~\ref{tbl:sa}. 
The {\em tolerance} for each case is set to have less than
5\% of the actual (row-wide) matches missed under the respective error rate,
according to Table~\ref{tbl:sa}.  In Fig.~\ref{fig:sa}, the X-axis captures the
overshoot with respect to the {\em tolerance};
the Y-axis,
the anticipated probability of occurrence of the corresponding {\em less similar
match}.
Since \arch's sensing threshold is variation-aware, the probability of 
{\em less similar match}es
featuring only an overshoot by one (base mismatch) becomes
predominant.  The probability of occurrence for 
{\em less similar match}es
with an
overshoot by 2 is still notable.  However, an overshoot by 3 or more is
negligible, with an expected probability of less than 0.3\% under both error
rates. 
As
Fig.~\ref{fig:sa} shows, the expected overshoot can be well bounded below 4
(base) mismatches by carefully setting the {\em tolerance}, where for more than
99\% of the 
cases,
the overshoot becomes only 1 or 2. The only problem with {\em less similar
matches}
(even if accompanied by a small overshoot) is the
uncertainty incurred in the mapping results. Luckily, {\em tolerance} is already
a fuzzy concept for {\em read mapping}.  On the other hand, the variation-aware
tuning of the SA sensing threshold (Section~\ref{sec:tol})  
prevents the interpretation of actual matches as mismatches. 
}

\subsection{Input Dataset}
\label{sec:genome}
\noindent We use a real human genome, \texttt{g1k\_v37}, from the 1000 genomes
project~\cite{genome} as the \srg\ genome; 
and 
20 million 100-base long 
real {\em reads} from NA12878~\cite{NA12878} as a {\em query} dataset.
For mapping accuracy analysis, we further
generate 20 million more 
{\ssq}s using 100-base long randomly picked sub-sequences from this \srg, which
we corrupt considering
a {read} error probability
of 0.1\%  (to mimic modern Illumina platforms), 
a single substitution~\footnote{i.e., single-nucleotide polymorphism, SNP, the dominant
type of substitutions induced by genomic variations}
probability of 0.09\%, and a short indel probability of 0.009\%~\cite{}.
For a fair comparison (not to favor \arch) we choose
the number of {\ssq}s to have the {\em reference} + {\em query}s fit into the
main memory of the GPU, such that the GPU does not suffer from extra
energy-hungry data communication.  We also limit the evaluation to a single
\arch\ card to keep the resource utilization comparable to the baselines.

\subsection{Baseline for Comparison}
\label{sec:base}
\noindent As a comparison baselines, we pick a highly optimized GPU
implementation of the popular BWA algorithm, SOAP3-dp~\cite{luo2013soap3},
and a short-read aligner hardware accelerater, GenAx~\cite{genax}.  A pure
software-based implementation of \arch\ is orders of magnitude slower than
SOAP3-dp.
%
We evaluate the throughput performance and
power consumption of SOAP3-dp on an NVIDIA Tesla K40 GPU.
We measure the power consumption of the GPU using NVIDIA-SMI
(System Management Interface) command.  We use the 
same \srg\ and \ssq\ dataset (Section~\ref{sec:genome}) as
\arch\ as inputs.  
We compare \arch\ against two different configurations of SOAP3-dp:
The first one, SOAP$_{SUB}$, only handles substitutions;
the second one, SOAP$_{ALL}$, captures all prevalent manifestations of {read}
errors and genomic variations.

\subsection{Design Space Exploration}
\label{sec:dse}
We sweep $M$, $C$, $N$, and $L$ in the evaluation section
to find the highest
throughput design, $\arch_{Perf}$, the most energy-efficient
design, $\arch_{Energy}$, and a design optimized for both,
$\arch_{Optim}$. We sweep $M$ from 16GB to 128GB, and $L$
from 10 to 15. 
For each $L$, we find the maximum $N$ where PMIT size and 
$N\times$PMITIL size fits in the given $M$ budget.
Besides,
we cap $N$ at 512, to limit $C$, number of LPDDR4-4266 channels
needed to feed PMIs to FilterUs, at 64.
Fig~\ref{fig:setupsweep} depicts how the ranges of area and power
consumption of \arch\ looks like for different evaluated designs
with different $M$, $C$, $N$, and $L$ parameters.

\begin{figure}[tp]
  \begin{center}
  \includegraphics[width=0.85\columnwidth]{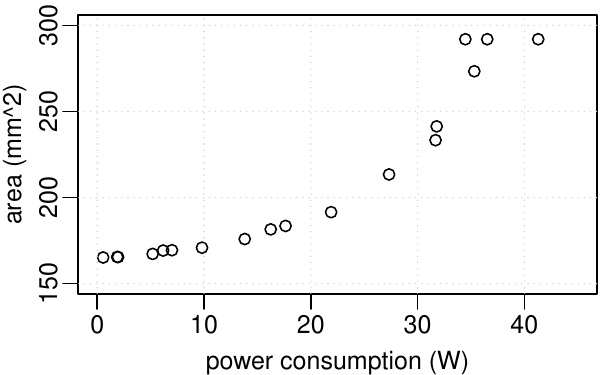}
  \caption{\arch's area and power consumption for different evaluated designs.
  \label{fig:setupsweep}}
\end{center}
\vshrink{0.85}
\end{figure}

\ignore{
\redHL{[This part will be revisited after I finish the evaluation.]}

\noindent We evaluate three \arch\ configurations: The first one,
{\arch$_{SUB}$} only implements Phase 1 and Phase 2 from
Fig.~\ref{fig:phase}, hence covers
substitutions and reverse complements. 
The second one,
{\arch$_{ALL}$}, implements Phases 1, 2, and 3, hence can capture {\em
prefix} corruptions and short indels, as well. Finally,
{\arch++} represents the implementation where we pair {\arch$_{ALL}$} with SOAP3-dp,
as detailed in Sect.~\ref{sec:hier}. 
}

\ignore{
specifically, we feed  every 500K missed {\em query}s (after Phase 3) 
periodically 
(after processing every 500K {\ssq}s)
to achieve the best possible accuracy
's unaligned reads,
\redHL{
as well as aligned ones along with their final matching indices,
}
are fed to SOAP3-dp 
periodically (after processing every 500K {\ssq}s)
to achieve the best possible accuracy
\redHL{
and generate SAM format output.
}
Fig.~\ref{fig:setupsweep} 

\redHL{
[Probably figures or tables here to quantify area and power for different 
(M,C,L,N) tuples. I will update this after I am done with all evaluations.]
}
}

\section{Evaluation}
\label{sec:eval}



\subsection{Throughput Performance and Energy}
\label{sec:et}

\noindent Larger
\seed\ values prune the search space more, resulting in a progressively lower
number of search operations in processing each \ssq.
Increasing $L$ from 10 to 15 decreases average number of search operations
per query from 30.2K to 4.6K. On the other hand for a given $M$ budget,
lower $L$ leads to larger $N$ values, which consequently increases parallelism
level of \arch. Therefore, we need to sweep $L$ for different $M$ DRAM budgets
to find the optimum $L$ value.

\begin{figure*}[tp]
\begin{center}
  \vshrink{0.2}
  \subfloat[Throughput Performance]{  
	\includegraphics[height=4.6cm]{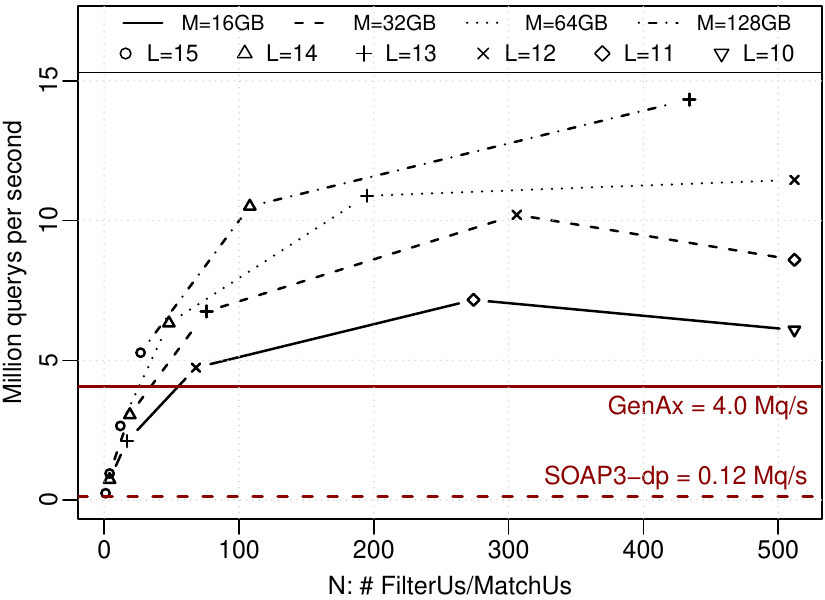}
		\label{fig:throughput}
 } 
 \subfloat[Energy]{  
	\includegraphics[height=4.6cm]{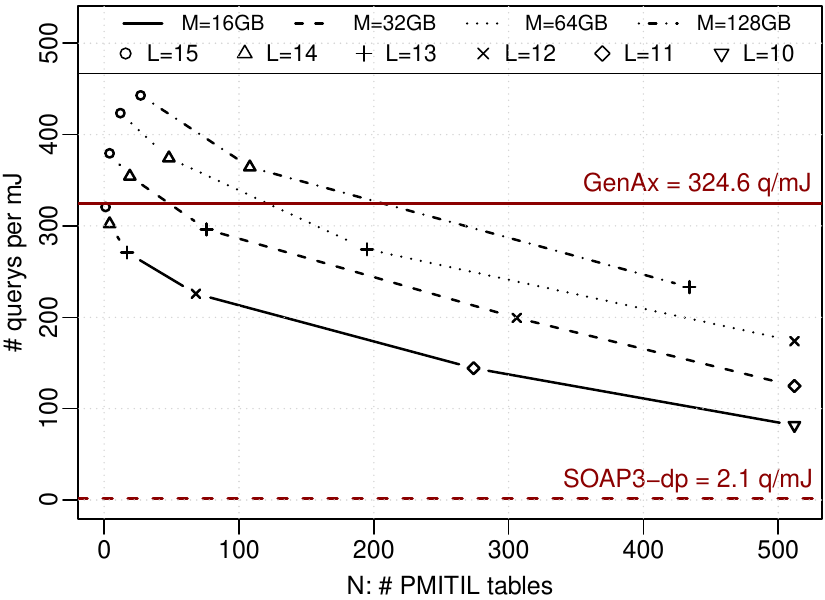}
		\label{fig:energy}
  }
 \subfloat[Trade-off]{  
	\includegraphics[height=4.6cm]{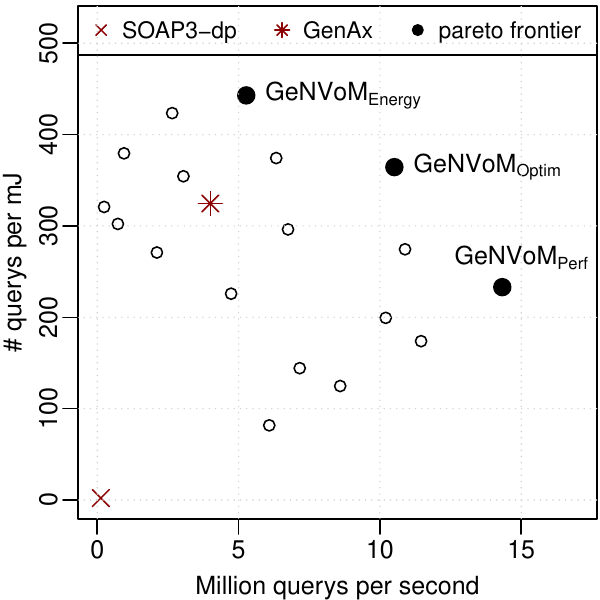}
		\label{fig:trade}
  }
  \vshrink{0.1}
  \caption{Throughput performance and energy consumption.
  \label{fig:eval}}
\end{center}
\vshrink{0.4}
\end{figure*}

Fig.~\ref{fig:throughput} depicts the throughput performance of \arch.
Y-axis represents the number of {\ssq}s mapped per second.
X-axis shows the number of FilterUs/MatchUs used, $N$. 
The two red horizontal lines correspond to the throughput of the two baselines, 
SOAP3-dp and GenAx.
Each line corresponds to a different $M$ budget. Each point belongs to a different
$L$. We do not include the $L$ values which violate the cap of $N$, 512.

As Fig.~\ref{fig:throughput} shows, starting from $L$ of 15 (on the left side),
decreasing $L$ increases $N$, and consequently increases throughput. However,
as $L$ goes lower than 12, the extra number of searches dominates the benefits of
a larger $N$, leading to slow-down in performance. Besides, we see that having a
larger $M$ budget improves the overall throughput. While most points are above both
baseline lines, the fastest design, $\arch_{Perf}$, gives up to 113.5$\times$ 
speed-up over SOAP3-dp, and 3.6$\times$ over GenAx, for $M=128GB$, $L=13$, $C=55$, and $N=434$.
$\arch_{Perf}$ consumes 35.3 Watts of power on average, and 273.3 $mm^2$ of area.

Fig.~\ref{fig:energy} demonstrates the energy efficiency of \arch.
Y-axis represents the number of {\ssq}s mapped per Millie Joule of energy (q/mJ).
X-axis shows the number of FilterUs/MatchUs used, $N$. 
The two red lines correspond to the q/mJ rate of the two baselines.
Similar to the throughput figure, each line represents a different $M$ budget, and
each point shows a different $L$. 

As we see in Fig.~\ref{fig:energy}, energy efficiency worsens for smaller $L$s,
since number of searches increases for smaller $L$s, and at the same time, $N$
grows and consequently we see more power consumption in FilterUs, MatchUs, and
NOCs. \arch outperforms SOAP3-dp in energy-efficiency for all evaluated parameters,
while still many designs outperform GenAx. The most energy efficient design, 
$\arch_{Energy}$, mapping 442.9 queries per Millie Joule, corresponds to $M=128GB$,
$L=15$, $C=4$, and $N=27$. $\arch_{Energy}$ outperforms energy-efficiency of 
SOAP3-dp by 210.9$\times$ and GenAx by 1.36$\times$, while consuming 5.2 Watts of
power and 167.3 $mm^2$ of area.

Finally, Fig.~\ref{fig:trade} depicts the trade-off space of throughput (x-axis)
versus energy-efficiency (y-axis) for different parameters. Points closer to the
top-right corner are faster and more energy efficient. The two red points represents
where baselines are. All points outperform SOAP3-dp in both throughput and 
energy-efficiency, while a few can be both faster and more energy-efficient
compared to GenAx. Besides, we see the pareto frontier points marked with black.
We pick the third design optimized for both performance and energy-efficiency,
$\arch_{Optim}$, from the pareto-frontier points, with $M=128GB$, $L=14$, $C=14$,
and $N=108$. 
$\arch_{Optim}$, consuming
21.9 Watts of power and 191.5 $mm^2$ of area, maps 10.5 million querys per second
(84.0$\times$ faster than SOAP3-dp and 2.6$\times$ faster than GenAx) and 364.4
querys per Millie Joule (173.5$\times$ better than SOAP3-dp and 1.12$\times$ better
than GenAx). All three pareto-frontier designs correspond to $M=128GB$, which shows
\arch benefits from larger memory space in both throughput and energy-efficiency.

Without loss of generality for $L=15$, \arch
maps
45.4\% of the {\em queries} in Phase 1;
42.1\% in Phase 2 (reverse complement);
and 8.4\%, in Phase 3. 
Specifically, in Phase 3, \arch 
maps
3.0\% of the {\em queries} after anchoring the first half; 
2.6\% after anchoring the second half;
1.8\% after anchoring reverse complement of the first half;
and 1.0\% after anchoring reverse complement of the second half, respectively.
This renders a mapping rate (i.e., the share of successfully mapped {\ssq}s over all)
of around 96.0\% for \arch, while SOAP3-dp can
map 97.4\% of the {\ssq}s.

\ignore{
\subsection{System Bottlenecks}
\label{sec:bott}
\noindent We next identify system bottlenecks.  According to simulation results,
MatchU is the most time consuming stage of the {\arch$_{ALL}$} pipeline.
Fig.~\ref{fig:piedelay} depicts the share of time spent in each unit of the
MatchU, for 16 chips, which represents the fastest design point.
The throughput bottleneck is communication of {\ssq}s
and PMIs, taking 57.9\% of the time.  In Fig.~\ref{fig:piedelay}, {\em Logic}
refers to DispatchU, queues, and controllers in MatchU; {\em Array}, to the TCAM
arrays.  24.8\% of the time goes to actual search operations in the TCAM
arrays. Besides, 17.3\% of the time is spent in DispatchU (and the rest of the logic).

\begin{figure}[htp]
\vshrink{0.4}
\begin{center}
  \includegraphics[height=0.5cm]{FIG/legend.pdf} \\
  \vshrink{0.1}
  \subfloat[Time]{  
	\includegraphics[height=3.3cm]{FIG/isca18/isca18_pie_delay.pdf}
		\label{fig:piedelay}
 } 
 \subfloat[Energy]{  
	\includegraphics[height=3.3cm]{FIG/isca18/isca18_pie_energy.pdf}
		\label{fig:pieenergy}
  }
  \vshrink{0.1}
  \caption{Time and energy break-down. 
  \label{fig:discuss}}
\end{center}
\vshrink{0.5}
\end{figure}

Fig.~\ref{fig:pieenergy} depicts the energy share of each {\arch$_{ALL}$} unit, for the
most energy efficient design point, using 16 chips.  {\em Logic} refers to the total
energy consumption of all logic units incorporated in FilterU and MatchU; {\em
DRAM}, to the PMI tables; {\em Array}, to the TCAM arrays. Since {\arch$_{ALL}$} keeps 16
chips utilized in parallel, most of the energy goes to communication, rendering
a share of 46.2\%. Performing actual search operations in TCAM arrays consumes
23.7\% of the energy. 19.7\% of the energy goes to search space pruning (i.e., accessing the PMI tables); and 10.4\%, to logic operations.
}

\subsection{Mapping Accuracy}
\label{sec:acc}
\noindent Since we were not able to run experiments using GenAx,
we compare the accuracy of \arch\ to SOAP3-dp only. Besides,
SOAP3-dp outperforms accuracy of BWA~\cite{luo2013soap3}, which
has very similar accuracy to GenAx. Therefore, SOAP3-dp is expected
to have better accuracy compared to GenAx.
To compare the mapping accuracy of \arch\ to SOAP3-dp, 
we use a simulated input dataset with known
expected matching indices.
We differentiate between two cases: 
(1) the {\em query} is aligned to a wrong portion of the {\em reference};
or (2) the {\em query} is not aligned to any portion of the {\em reference}.
Table~\ref{tbl:accur} shows the corresponding rate of occurrence for (1)
as the {\em Misalignment rate}; for (2), as
the {\em Miss rate}, considering different configurations. 
We should note that misaligned {\ssq}s are still mapped to a portion of the {\em
reference}, which may be similar enough. Therefore, {\em misalignment rate} does
not necessarily correspond to an error rate.
The numbers in Table~\ref{tbl:accur} reflect all errors caused by \arch's
prefix corruption,
indel mishandling, and SA's false positives.

\begin{table}[h]
  \resizebox{\linewidth}{!}{
\centering
\begin{tabular}{c|cccc|}
\cline{2-5}
                                                      & \textbf{\arch$_{Energy}$} &
\textbf{\arch$_{Optim}$} & \textbf{\arch$_{Perf}$} & \textbf{SOAP3-dp} \\ \hline
\multicolumn{1}{|c|}{\textbf{Misalignment Rate}}      & 2.94\%                 & 2.97\%                & 2.99\%         & 1.12\%               \\
\multicolumn{1}{|c|}{\textbf{Miss Rate}} & 0.03\%                 & 0.03\%                & 0.04\%         & 0.01\%               \\ \hline
\multicolumn{1}{|c|}{\textbf{Total}} & 2.97\%                 & 3.0\%                & 3.03\%         & 1.13\%               \\ \hline
\end{tabular}
}
\caption{Mapping accuracy of \arch\ w.r.t. SOAP \label{tbl:accur}	}
\vshrink{0.2}
\end{table}


We observe that both {\em misalignment rate} and {\em miss rate} are slightly 
increasing for larger $L$ values (from \arch$_{Energy}$ to \arch$_{Perf}$), due
to higher prefix corruption probability. 
However, the overall accuracy of different \arch\ designs considering different 
$L$ values stay in a similar range.
Generally, \arch fails to align only
around 0.03\% more querys compared to SOAP3-dp, while misaligning around 1.85\%
more querys.

Next, we evaluate the effectiveness of \arch's anchoring phase (phase 3 from
Section~\ref{sec:hier}), in improving the accuracy of mapping after the first
2 phase. Without loss of generality, \arch$_{Energy}$ fails to map 2.88\% of the 
{\ssq}s, after the first 2 phases, mapping normally and the reverse complement. 
Phase 3 of \arch$_{Energy}$, featuring anchoring on the other hand, 
can map 98.6\% of the {\ssq}s that the first two phases
misses
(due to indels and prefix corruption), 
while aligning 91.2\% of them to the correct index of the reference genome.
This analysis demonstrates how effective multi-phase search is in improving the
mapping accuracy when dealing with querys failed-to-map by the first two phases. 

SOAP3-dp let user trade-off accuracy for higher throughput. In this mode,
SOAP3-dp does not map using the dynamic programming kernel, and only relies
on capturing substitution errors. User can set the accuracy level of the program
by tuning the number of acceptable mismatches, which is similar to the {\em tolerance}
level of \arch. Lower {\em tolerance} level leads to lower accuracy. At the
same time, lower {\em tolerance} improves throughput of SOAP3-dp, since it
performs lower number of search operations. On the other hand, different
{\em tolerance} values does not affect \arch's throughput noticeably, since
neither FilterU's or MatchU's performance depend on it. More specifically,
\arch\ can tune {\em tolerance} by only adjusting SA's threshold, which
negligibly affects TCAM search latency only.

\begin{figure}[tp]
  \begin{center}
  \includegraphics[width=0.8\columnwidth]{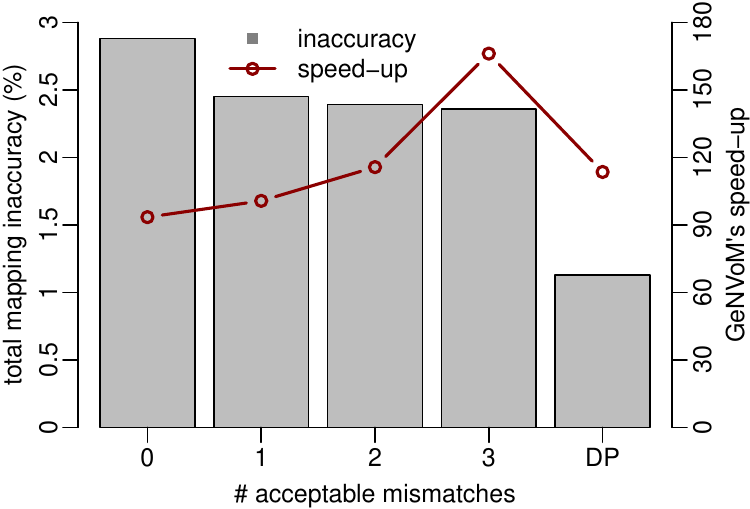}
  \caption{Impact of {\em tolerance} on SOAP3-dp's accuracy and peformance.
  \label{fig:sweepaccur}}
\end{center}
\vshrink{1}
\end{figure}

Fig~\ref{fig:sweepaccur} depicts how SOAP3-dp's accuracy (left y-axis) changes
for different number of acceptable mismatches (x-axis). The right-most bar
belongs to default SOAP3-dp parameters, which includes the dynamic programming
kernel as well. The line in the figure corresponds to the speed-up of \arch$_{Perf}$
versus SOAP3-dp (right y-axis). As we see, for smaller {\em tolerance}s,
SOAP3-dp becomes less accurate, and at the same time faster, which leads to
lower \arch's speed-up. The lowest accuracy of SOAP3-dp, 2.88\% for a
{\em tolerance} of 0, is close to accuracy of \arch$_{Perf}$. However, being
21.5\% faster compared to the default SOAP3-dp, it is still 93.4$\times$ slower
compared to \arch$_{Perf}$. Therefore, \arch\ outperforms even an iso-accuracy baseline 
by two orders of magnitude.

\noindent{\bf A note on ``acceptability'':} 
In a typical NGS {\em query} dataset, 
all {\em query}s, if concatenated back to back, would be at least 
$50\times$~\cite{ajay} longer than the 
{\em reference} genome. Therefore, even if we miss the mapping of a few percent
of the {\em query}s (due to 
different {\em read} errors/genome variations), we still have plenty of {\em
query}s to cover such missed regions of the {\em reference} -- which would be covered by the
missed {\em queries} were there no {\em read} errors/genome variations.
The average number of {\em query}s covering any given base in the
{\em reference} genome is called the {\em depth}.

Only missing {all} {\em query}s covering a specific base of
the {\em reference} would be problematic. 
What is the probability for \arch\ to miss all {\em query}s covering a given
base of
the {\em reference}?
Let the number of {\em query}s
covering a given base be $Q$, and let $min(Q)$ denote its minimum. 
$Q$
follows a Poisson distribution~\cite{lander}. 
For a representative {\em depth}
of 50, the probability of having a base covered by less than
10 {\em query}s is
$5.2\times 10^{-12}$, 
practically negligible.
Therefore, we can assume that all bases are covered at least by 10 {\em query}s,
i.e., that $min(Q)=10$.
The probability of \arch\ missing a mapping,
$P_{miss}$ is around $3.03\%$ 
(Table~\ref{tbl:accur}). Hence, the probability of \arch\ missing all {\em query}s
covering a base of the {\em reference} becomes 
${P_{miss}}^{min(Q)}=6.5\times 10^{-16}$,
  which is practically negligible.

\ignore{
\noindent{\bf Mismatches in fragmented matches vs. {\em tolerance}:} In handling
fragmented matches (Fig.~\ref{fig:match}), \arch\ compares each input \ssq\ to
two consecutive rows of the TCAM array. In the current implementation, both
steps of the search per row are subject to the same {\em tolerance} value.
Therefore, if a match is the case in both of the rows, while the number of base
mismatches per row remains below the {\em tolerance} (in the absence of false
positives), there is no guarantee that the cumulative number of base mismatches
(over the two consecutive rows) remains below the {\em tolerance}.  In fact, the
cumulative number may well exceed the {\em tolerance}.  As discussed in
Section~\ref{sec:sensest}, this would not necessarily lead to an error, but
rather, a less similar match, which indeed may be an actual match.  Let us
revisit the example from Section~\ref{sec:sensest}, where the NGS error rate =
0.1\% and {\em tolerance} = 1 (base mismatch). In this case, a fragmented match
would correspond to a mapping as if the {\em tolerance} was 2, which is likely
to pass as {\em accurate enough}.
}



\ignore{
\subsection{Sensitivity to tolerance}
\label{sec:scale}
\ignore{
\noindent 
In this study, we evaluate a single \arch\ DIMM. Simulation results show that
for the highest throughput design (16 chips), DRAM bandwidth utilization is
  17.7\%.  DDR4 technology can support more than one \arch\
  DIMM, based on this utilization profile, depending on the contention of the
  memory bus as a shared resource.  Multiple \arch\ DIMMs would not need extra
  DRAM space, since they can share the PMI tables if used for mapping to the
  very same \srg\ genome. However, each DIMM would get a different \ssq\ dataset
  to map.  Multi \arch\ DIMM support becomes more likely for a higher-bandwidth
  DRAM technology. 

\noindent We use {\ssq}s of 100-bases long throughout the evaluation. However, depending
on the NGS platform deployed, {\ssq}s can be longer or shorter.  {Longer {\ssq}s
(still within Illumina-like NGS platform boundaries) increase statistical
confidence.} \arch\ can practically support a wide range of \ssq\ lengths.  To
use a different length, following the methodology discussed in
Section~\ref{sec:sensest}, only the {\em tolerance} and the sense amplifier
threshold should be tuned accordingly.  No further changes are necessary, as
long as the \ssq\ length stays within the row length of the TCAM arrays, since
\arch\ performs search at row granularity.

Fig.~\ref{fig:inps} depicts how the throughput of \arch\ (for \seed=15) and
Barracuda scale with different input {\em query} lengths, as captured by the
X-axis.  Left Y-axis denotes throughput in terms of total number of {\em query}s
processed per second; right Y-axis, \arch's relative throughput improvement over
Barracuda.  As Barracuda cannot handle {\em query}s longer than 150 bases, we
had to limit the maximum length to 150.  On the other hand, \arch\ can handle
longer {\em query}s, as long as each {\em query} fits in a single row of the
TCAM arrays.  According to Fig~\ref{fig:inps}, \arch's throughput decreases with
increasing {\em query} length, since communication takes proportionally more
time.  However, \arch\ still is more scalable, and the relative throughput
improvement over Barracuda increases with the {\em query} length. For the
longest {\em query} length that Barracuda can handle, \arch\ provides
$14.1\times$ higher throughput.
}

\noindent
{A higher {\em tolerance} value can effectively decrease the rate of missed
{\ssq}s for mapping, which have a larger number of their bases corrupted (due to
{\em read} errors and genomic variations) than
{\em tolerance}.
Fig.~\ref{fig:sensperf}
depicts how the throughput of \arch$_{SUB}$ (with 16 chips) and 
SOAP$_{SUB}$ scale with
different {\em tolerance} values captured by the X-axis. Recall that  
{\em tolerance} indicates the number of base mismatches that are
permitted to pass as matches during mapping.
Left
Y-axis denotes throughput in terms of total number of {\em query}s processed per
second; right Y-axis, relative throughput improvement of \arch$_{SUB}$ 
over SOAP$_{SUB}$.

For \arch$_{SUB}$, changing {\em tolerance} is equivalent to changing sense
amplifier (SA) threshold
voltage, following the methodology from Section~\ref{sec:sensest}.  To increase
{\em tolerance}, SA threshold voltage should decrease, which marginally
slows the SA down. 
On the other hand, increasing {\em tolerance} increases the mapping rate (i.e.,
1-{\em Miss Rate} per Table~\ref{tbl:accur}) of \arch$_{SUB}$,
which leads to a lower number of {\ssq}s going through extra search phases
(Fig.~\ref{fig:phase}).
Therefore, increasing {\em
tolerance} increases \arch$_{SUB}$'s throughput marginally (Fig.~\ref{fig:sensperf}).

SOAP$_{SUB}$'s throughput, however, degrades
notably with increasing {\em tolerance}, {since it cannot prune the
search space as aggressively under a higher {\em tolerance} value.} 
Consequently, \arch$_{SUB}$'s throughput improvement over SOAP$_{SUB}$
increases for higher {\em tolerance} values (Fig.~\ref{fig:sensperf}). For a {\em tolerance} of 4 (largest allowed in SOAP$_{SUB}$), 
\arch$_{SUB}$'s throughput improvement over SOAP$_{SUB}$ becomes $19.3\times$.

Fig.~\ref{fig:sensaccept} compares mapping rate (Y-axis) of \arch$_{SUB}$ and
SOAP$_{SUB}$ under different {\em tolerance} values (X-axis). We observe that 
\arch$_{SUB}$'s mapping rate closely tracks SOAP$_{SUB}$'s. For a {\em tolerance} of 4,
\arch$_{SUB}$ misses only around 2.9\% more {\ssq}s, when compared to SOAP$_{SUB}$.
To summarize, \arch$_{SUB}$ can handle higher {\em tolerance} values when compared to SOAP$_{SUB}$,
with negligible decrease in mapping rate.

\begin{figure}[tp]
\vshrink{0.1}
\begin{center}
\hspace{0.1cm}
\subfloat[throughput]{  
	\includegraphics[height=4.1cm]{FIG/isca18/isca18_sens_accept.pdf}
		\label{fig:sensperf}
 } 
 \hspace{.1cm}
 \subfloat[mapping rate]{  
	\includegraphics[height=4.1cm]{FIG/isca18/isca18_sens_accept_rate.pdf}
		\label{fig:sensaccept}
  }
    \caption{
	  Sensitivity to {\em tolerance}.
  \label{fig:scal}}
\end{center}
\vshrink{0.8}
\end{figure}
}
\ignore{
\subsection{Extension to Similar Application Domains} \noindent While tailored
to {\em read mapping}, \arch\ is fundamentally applicable to similar
search-intensive problems.  For such extensions, we first need to determine the
minimum number of bits needed to represent each character in the alphabet of the
new problem, as covered in Section~\ref{sec:encode}, and an acceptable {\em
tolerance} level.  As the number of bits representing each character grows, the
number of bit-mismatches per character-mismatch grows, as well.  Consequently,
sensing becomes even easier to design.  We next need to tune the {\em threshold}
of the sense amplifiers (Section~\ref{sec:sensest}), and pick a \seed\ to
maximize the energy efficiency and throughput.  We leave such exploration to
future work.
}

\section{Related Work}
\label{sec:rel}
{\noindent \bf (Short) Read Mapping:} 
With no pre-processing of the {\em reference}, the computational complexity 
scales (at least) linearly with the {\em reference} length.
Therefore, 
popular software implementations such as
SOAP~\cite{li2008soap}, Eland (part of the Illumina suite), and
MAQ~\cite{li2008mapping}
adapt hash-based pre-processing. 
SOAP2~\cite{li2009soap2},
Bowtie(2)~\cite{langmead2009ultrafast,langmead2012fast}, 
BWA~\cite{li2009fast,li2010fast}, on the other hand, use 
the (more memory efficient) Burrows-Wheeler
Transformation (BWT)
of the {\em reference}.
One baseline for comparison, SOAP3-dp~\cite{luo2013soap3}, represents an
open-source 
GPU implementation of BWA, which
can also
handle noise in {\em read}s, however,
unlike \arch, the computational complexity depends on the {\em
tolerance} value.
Genax~\cite{genax}, the other baseline for comparison, is an
automata-based accelerator. Similar to SOAP3-dp, computational
and space complexity of GenAx depend on the {\em tolerance} value,
unlike \arch.
%
High-throughput FPGA implementations~\cite{fpga1} 
%
also exist, at the expense of orders of magnitude
higher power consumption than \arch.
%
The exotic race
logic based dynamic programming accelerator~\cite{madhavan2014race} can find the similarity between two
strings corresponding to the {\em read} {and a sub-sequence of the} {\em
reference} in approx. 120ns while consuming 1nJ.  This is much
slower than \arch, and energy grows with the third power of {\em read}
length which can impair scalability.

Another study introduced an efficient filtering step for hash-based {\em
read mapping} using 3D-stacked memories~\cite{kimgenome}. This covers
filtering (i.e., search space pruning) only, which has a similar functionality
to \arch's FilterU.  
While the high
cost of quadratic-time dynamic programming algorithms for string (i.e., {\em
query}) matching motivated this work, \arch\ relies on
much faster and more energy-efficient string matching enabled by resistive TCAM.
Hence, \arch\ employs a simpler, low-latency filtering (incorporated
in FilterU), which is tailored to its more efficient string matching
(incorporated in MatchU).  Although higher-overhead
methods like ~\cite{kimgenome} can prune the search space further, embedding
such into \arch's FilterU would be overkill, rendering FilterU the
system bottleneck and diminishing the benefits from MatchU's fast and
energy-efficient similarity matching.  We believe that proposals
like~\cite{kimgenome} are more suitable for {\em long read mapping}, which
represents a different problem.
Besides
GenAx~\cite{genax}, recent work also includes a BWA-based hardware
accelerator featuring NVM~\cite{aligner}, which is significantly slower than \arch; and 
very effective hardware acceleration
for other significant bioinformatics algorithms~\cite{kevin_bio}.

{\noindent \bf {Resistive CAM Accelerators:}} Guo et al.~\cite{ipekTCAM,ipekACC}
explore 
TCAM for accelerating data-intensive
applications. Yavits et al.~\cite{yavits2015resistive} propose an associative
processor, which employs resistive CAM based look-up tables to implement diverse functions.
Kaplan et al.~\cite{kaplanSW} demonstrate how to 
efficiently implement Smith-Waterman algorithm (which tries to align strings of
similar length)
using resistive CAM based look-up tables.
%
Not being able to handle similarity matching under noise, neither of these are directly applicable to {\em read mapping}.
\ignore{
While \arch\ features non-volatile TCAM as~\cite{ipekTCAM, ipekACC} do,
\arch\ arrays do not include any of the priority index logic, population count
logic, or the reduction network from~\cite{ipekTCAM} or programmable
microcontrollers of~\cite{ipekACC}.
Instead, \arch\ tailors the data and control paths to {\em read mapping}, which
entails minimal logic for filtering, scheduling, and queuing search requests.
At the same time, \arch's TCAM arrays feature a novel similarity match
mechanism cut for {\em read mapping}. Neither of~\cite{ipekTCAM, ipekACC} are
directly applicable to {\em read mapping}.
}
Approximate resistive CAM is also proposed, either using exotic
cells~\cite{rahimi2} (which complicates match detection, and thereby restricts the row length
to at most 8 bits) or limiting the row length to 4 bits only (to find the
Hamming distance, their similarity metric,
robustly)~\cite{rahimihyperdimensional}.
\ignore{
The first design~\cite{rahimi2} relies on an exotic cell, which
connects $R_{low}$ ($R_{high}$) to ML on a (mis)match; the opposite
convention to the state-of-the-art per 
Fig.~\ref{fig:tcamCell}.
This complicates
match detection, thereby restricts the row length to at most 8 bits.
The second design~\cite{rahimihyperdimensional} 
does not use an exotic cell,
however, to find the Hamming distance (their similarity metric) robustly, limits
the row size to 4 bits only.
%
}
While these show great potential, restricted row
length hinders applicability to 
{\em read mapping} 
where longer 
(short) 
{\em read}s
are emerging with NGS improvements and where TCAM search happens at row-granularity.  \arch\ adds
support for approximate matches much less intrusively, by carefully tuning the
sense amplifier (SA) reference voltage in a variation-aware manner, without restricting the row
size.
%
No CAM array capable of only approximate matching would be sufficient to
implement an efficient {\em read mapping} accelerator by itself, as
demonstrated in Section~\ref{sec:motiv}.  

Recent representative demonstrations of resistive
TCAM include a 1Gbit PCM-based CAM from IBM
using IBM 90nm
technology~\cite{ibm_tcam} with a measured search latency of 1.9ns, and
two novel spintronic designs in 45nm~\cite{tcam1},
where a search 
takes approx. 0.6ns in 256-wide rows.
\arch\ can adapt any resistive CAM array, including these more recent proposals.
%

\ignore{
{\noindent \bf \hlnew{Promising Alternative Computing Paradigms:}}
The evaluated proof-of-concept implementation represents one  feasible point in the
rich design space of \arch. The system interface can take a different form, as
well.
3D stacking is an option, for example, to enable even more
parallelism subject to thermal budget. 
\ignore{
That said, the evaluated design point features TCAM which suits
very well to efficient similarity search 
and the evaluated interface leads to a relatively less intrusive design.
}
We could implement \arch\ using emerging in/near-memory logic, but the scale of
the problem would demand careful optimization for data communication between the
memory modules and logic embedded in/near memory.  }
%
\ignore{
Recently, Kim et al. have explored an efficient filtering step for hash-based
{\em read mapping} using 3D-stacked memories~\cite{kimgenome}.  This study
covers filtering (i.e., search space pruning) only, which has a similar
functionality to \arch's FilterU.  In line with our observations, the
authors are after pruning the search space in order to maximize
the throughput. 
%
While the high cost of ``quadratic-time dynamic programming'' algorithms for
string (i.e., {\em query})
matching 
motivated Kim et al.~\cite{kimgenome}, \arch\ already relies on 
much faster and more energy-efficient string matching enabled by 
TCAM search. Hence, \arch\ employs a simpler, low-latency filtering mechanism
(incorporated in the Filter Unit), which is tailored to this more efficient string
matching mechanism (incorporated in the Match Unit).
Although complex higher-overhead
methods like ~\cite{kimgenome} can prune the search space further, embedding
such into
\arch's Filter Unit would be overkill, rendering
Filter Unit the system bottleneck and
diminishing the benefits from Match Unit's fast energy-efficient similarity matching. 
We believe that proposals like~\cite{kimgenome} are more suitable for long {\em
read} mapping.
}

\section{Discussion \& Conclusion}
\label{sec:conc}
\ignore{
As semiconductor technology revolutionized computing, high-throughput
DNA sequencing technology 
revolutionized genomic research.  As a result, a progressively growing number of
short DNA sequences, generated at a faster rate than Moore's Law, needs to be
{\em mapped} to reference genomes which themselves represent full-fledged
assemblies of already sequenced DNA fragments.  This search-heavy data-intensive
mapping task does not need complex floating point arithmetic, and therefore, is
particularly suitable to in- or near-memory processing, where non-volatile
memory can accommodate the large memory footprint in an area and energy
efficient manner.  
}

\noindent A common critical first step in a diverse set of emerging bioinformatics
applications is {\em read mapping}, a search heavy, memory intensive approximate
pattern matching problem. This suggests TCAM-based acceleration, which by
definition features fast parallel in-memory search.  However, the
excessive energy consumption and lack of support for similarity matching under
noise hinders
the
direct application of TCAM-based search, irrespective of volatility,
where only non-volatile TCAM can accommodate the large memory footprint in an
area-efficient way.
This paper proposes \arch, a novel {\em read mapping} accelerator which unlocks
the throughput potential of non-volatile TCAM.
\arch\ {results in}
up to {113.5$\times$} ({3.6$\times$}) higher throughput while consuming up to {210.9$\times$} ({1.36$\times$}) less energy when
compared to a highly-optimized GPU (accelerator) implementation.

\ignore{
The evaluated proof-of-concept implementation represents one  feasible point in the
rich design space of \arch. The system interface can take a different form, as
well.
3D stacking is an option, for example, to enable even more
parallelism subject to thermal budget. 
That said, the evaluated design point features TCAM which suits
very well to efficient similarity search 
and the evaluated interface leads to a relatively less intrusive design.
This paper details the design of such a
near-(non-volatile)-memory sequence mapping accelerator, \arch, which results in
up to {113.5$\times$} ({3.6$\times$}) higher throughput while consuming up to {210.9$\times$} ({1.36$\times$}) less energy when
compared to a highly-optimized GPU (accelerator) implementation.
}

\ignore{
 Thereby \arch\ 
can improve 
the throughput 
by up to {113.5$\times$} ({3.6$\times$}); 
the energy consumption, by up to {210.9$\times$} ({1.36$\times$}), when compared to a
GPU (accelerator) baseline, which represents
one of the highest-throughput implementations known. 
}

Currently, short (i.e., 100-200 base long) {\em read}s from modern Illumina NGS
platforms~\cite{illumina} constitute more than 90\% of all {\em read}s in the
world.  This dominance is unlikely to quickly change in the near
future due to the progressively dropping sequencing cost of short {\em read}
technologies, rendering them significantly more cost-efficient than the long
{\em read} counterparts such as PacBio~\cite{pacBio} or Oxford
Nanopore~\cite{branton2008potential} (where {\em read} lengths can exceed tens
of thousands of bases).  The key benefit of long {\em read} sequencing
technologies comes from the capability of directly extracting long-range
information, and not necessarily from higher accuracy. 
That said, many emerging recent technologies such as 10xGENOMICS~\cite{10xG} can
obtain long-range information from short {\em read}s.  Although it is very hard
to predict the future exactly, considering practical facts such as market share
and market caps, we believe that short {\em read} platforms will remain
prevalent at least in the near future.  Accordingly, \arch\ is designed for
short {\em read mapping}.

Short {\em read mapping} and long {\em read mapping} represent two fundamentally different
problems. This is because of the difference in NGS and genomic variation induced
noise manifestations. Long {\em read}s are subject to more frequent and complex
corruptions, which gives rise to very different algorithms for the mapping
problem than short {\em read}s. For example, perturbation by a significant
number of indels is not uncommon~\cite{ross2013characterizing}.
A very efficient
hardware accelerator for long {\em read mapping} has also been proposed
recently~\cite{dally}. 
As such accelerators are highly optimized for long {\em read}s 
(which suffer from more complex noise manifestation), 
they are by construction sub-optimal for short {\em read mapping} 
(for which substitutions are dominant per Fig. 3).
%
%

The evaluated proof-of-concept implementation represents one  feasible point in the
rich design space of \arch. 
3D stacking is an option, for example, to enable even more
parallelism subject to a stringent thermal budget, where 
the scale of
the problem demands careful optimization for data communication between the
memory modules and logic embedded in/near memory, which we reserve for future
work.

\ignore{
can improve the throughput performance and energy efficiency of read mapping by
several orders of magnitude over the state-of-the-art, without compromising
mapping accuracy, which will likely pave the way for significantly quicker
discoveries to revolutionize medicine and biology. Solutions of the same nature
as \arch\ are critical to be able to keep up with the projected data generation
rates of NGS platforms.
}

\bibliographystyle{ieeetr}
\bibliography{main}

\end{document}